\RequirePackage{fix-cm}
\documentclass[twocolumn]{svjour3}          
\smartqed  
\usepackage[utf8]{inputenc}
\usepackage{graphicx}
\usepackage{booktabs}
\usepackage{color}
\usepackage{amsmath,amssymb,amsfonts}
\usepackage[hidelinks]{hyperref}

\usepackage{comment}
\usepackage{tikz}
\usetikzlibrary{automata, positioning, arrows}
\usepackage{color}
\usepackage{colortbl}
\usepackage{multirow}
\usepackage{balance}

\usepackage{longtable}
\usepackage{algorithm}
\usepackage[noend]{algpseudocode}
\algdef{SE}[SUBALG]{Indent}{EndIndent}{}{\algorithmicend\ }%
\algtext*{Indent}
\algtext*{EndIndent}

\usepackage{fancyvrb}
\usepackage{subfig}
\usepackage{paralist}

\usepackage{listings}
\definecolor{dkgreen}{rgb}{0,0.6,0}
\definecolor{gray}{rgb}{0.5,0.5,0.5}
\definecolor{mauve}{rgb}{0.58,0,0.82}
\newcommand\vldbdoi{XX.XX/XXX.XX}

\newcommand\vldbavailabilityurl{http://vldb.org/pvldb/format_vol14.html}

\usepackage{xspace}
\newcommand*{\eg}{e.g.,\xspace}
\newcommand*{\ie}{i.e.,\xspace}

\newcommand*{\etal}{\emph{et~al.}\xspace}
\newcommand*{\omem}{\textcolor{blue}{\bf{OM}}\xspace}

\newcommand{\insql}[1]{\textsf{#1}}

\renewcommand{\leq}{\leqslant}

\def\w#1{{\cellcolor[gray]{0.5}\color{white}\textsf{#1}}}

\newcommand{\word}[1]{w_{#1}}
\newcommand{\Node}{\mathit{Node}}
\newcommand{\vid}{\mathit{vid}}
\newcommand{\eid}{\mathit{eid}}
\newcommand{\lab}{\mathit{label}}
\newcommand{\Edge}{\mathit{Edge}}

\newcommand{\tim}{\mathit{time}}
\newcommand{\Xitab}{\mathit{Active}}

\newcommand{\States}{\mathit{States}}
\newcommand{\Extend}{\mathit{Extend}}

\newcommand{\pmatch}{partial-match\xspace}
\newcommand{\od}{on-demand\xspace}
\newcommand{\duck}{DuckDB\xspace}
\newcommand{\hyper}{HyPer\xspace}

\newcommand{\Hist}[2]{#1_{\leqslant #2}}

\newcommand{\xx}{\mathtt{c}}

\newcommand{\cust}{{\tt{cst}}}
\newcommand{\emp}{{\tt{emp}}}

\newcommand{\msg}{{\tt{msg}}}
\newcommand{\visit}{{\tt{visit}}}
\newcommand{\room}{{\tt{ofc}}}

\newcommand{\ta}{\mathit{TA}}

\newlength\myindent
\begin{document}

\title{Temporal graph patterns by timed automata}

\author{Amir Aghasadeghi         \and
        Jan Van den Bussche  \and 
       Julia Stoyanovich
}


\institute{A. Aghasadeghi 
\at New York University, USA\\
\email{amirpouya@nyu.edu}
\and 
J.Van den Bussche 
\at Hasselt University, Belgium\\
\email{jan.vandenbussche@uhasselt.be}
\and
J. Stoyanovich
\at New York University, USA\\
\email{stoyanovich@nyu.edu}
}

\maketitle
\begin{abstract}
Temporal graphs represent graph evolution over time, and have been receiving considerable research attention. Work on expressing temporal graph patterns or discovering temporal motifs typically assumes relatively simple temporal constraints, such as journeys or,  more generally, existential constraints, possibly with finite delays.  In this paper we propose to use timed automata to express temporal constraints, leading to a general and powerful notion of temporal basic graph pattern (BGP).  

The new difficulty is the evaluation of the temporal constraint on a large set of matchings.  An important benefit of timed automata is that they support an iterative state  assignment, which can be useful for early detection of matches and pruning of non-matches.  We introduce algorithms to retrieve all instances of a temporal BGP match in a graph, and present results of an extensive experimental evaluation, demonstrating interesting performance trade-offs. 
We show that an on-demand algorithm that processes total matchings incrementally over time is preferable when dealing with cyclic patterns on sparse graphs.  On acyclic patterns or dense graphs, and when connectivity of partial matchings can be guaranteed, the best performance is achieved by maintaining partial matchings over time and allowing automaton evaluation to be fully incremental. The code and datasets are available at \url{http://github.com/amirpouya/TABGP}.
\end{abstract}

\section{Introduction}
\label{sec:intro}


Graph pattern matching, the problem of finding instances of one
graph in a larger graph, has been extensively studied since the
1970s, and numerous algorithms have been
proposed~\cite{cheng2008fast,conte2004thirty,fan2010graph,lai2019distributed,reza2020approximate,ullmann1976algorithm}.  Initially, work in this area focused on
static graphs, in which information about changes in the graph is
not recorded.  Finding patterns in static graphs can be helpful
for many important tasks, such as finding mutual interests among
users in a social network.  However, understanding how interests
of social network users evolve over time, support for contact
tracing, and many other research questions and applications
require access to information about how a graph changes over
time.  Consequently, the focus of research has shifted to
pattern matching in \emph{temporal graphs} for tasks
such as finding temporal motifs, temporal journeys, and temporal
shortest paths
\cite{DBLP:conf/sigmod/GurukarRR15,DBLP:journals/corr/abs-1107-5646,DBLP:conf/wsdm/ParanjapeBL17,rost2022distributed,semertzidis2016durable,xu2017time,zfy-temporal-clique,DBLP:conf/edbt/ZufleREF18}.   

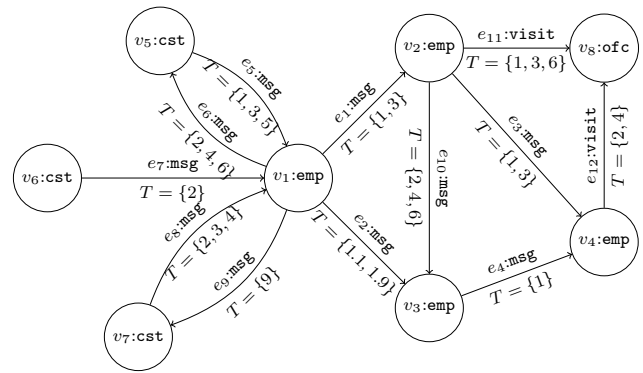
\begin{figure}
  \centering
   \resizebox{\columnwidth}{!}{
    \begin{tikzpicture}[on grid, node distance=90pt,auto]
      \node[state,minimum size= 10pt, node distance=90pt]   (v1)                     {$v_1$:\emp};
      \node[state,minimum size= 10pt, node distance=90pt]   (v2) [above right=of v1] {$v_2$:\emp};
      \node[state,minimum size= 10pt, node distance=90pt]   (v3) [below right=of v1] {$v_3$:\emp};
      \node[state,minimum size= 10pt, node distance=95pt]   (v5) [above left=of v1]   {$v_5$:\cust};
      \node[state,minimum size= 10pt, node distance=122pt]  (v6) [left=of v1]        {$v_6$:\cust};
      \node[state,minimum size= 10pt, node distance=110pt]   (v7) [below left=of v1]  {$v_7$:\cust};
      
      \node[state,minimum size= 10pt, node distance=85pt]   (v8) [right =of v2] {$v_8$:\room};
      \node[state,minimum size= 10pt, node distance=95pt]   (v4) [below =of v8] {$v_4$:\emp};

    \path[->] 
     (v1) edge  node [sloped,above] {$e_1$:\msg} node[sloped,below]{\small{$T=\{1,3\}$}} (v2)
     (v1) edge  node [sloped,above] {$e_2$:\msg} node[sloped,below]{\small{$T=\{1.1,1.9\}$}} (v3)
     (v2) edge  node [sloped,above] {$e_3$:\msg} node[sloped,below]{\small{$T=\{1,3\}$}} (v4)
     (v3) edge  node [sloped,above] {$e_4$:\msg} node[sloped,below]{\small{$T=\{1\}$}} (v4)
    
     (v5) edge [bend  left=25] node [sloped,above] {$e_5$:\msg} node[sloped,below]{\small{$T=\{1,3,5\}$}} (v1)
     (v1) edge [bend  left=25] node [sloped,above] {$e_6$:\msg} node[sloped,below]{\small{$T=\{2,4,6\}$}} (v5)
     
     (v2) edge node [sloped,above] {$e_{10}$:\msg} node[sloped,below]{\small{$T=\{2,4,6\}$}} (v3)
     
     (v6) edge  node [sloped,above] {$e_7$:\msg} node[sloped,below]{\small{$T=\{2\}$}} (v1)

     (v7) edge [bend  left=25] node [sloped,above] {$e_8$:\msg} node[sloped,below]{\small{$T=\{2,3,4\}$}} (v1)
     (v1) edge [bend  left=25] node [sloped,above] {$e_9$:\msg} node[sloped,below]{\small{$T=\{9\}$}} (v7)
     
     (v2) edge  node [sloped,above] {$e_{11}$:\visit} node[sloped,below]{\small{$T=\{1,3,6\}$}} (v8)
     (v4) edge  node [sloped,above] {$e_{12}$:\visit} node[sloped,below]{\small{$T=\{2,4\}$}} (v8)

    ;
    \end{tikzpicture}
    }
  \caption{Example of a temporal graph.
Each edge is associated with the set of timepoints during which
  it is active.}
  \label{fig:def2}
\end{figure}

Figure~\ref{fig:def2} gives our running example, showing an
interaction graph, where each node represents an
\emph{employee} (node label \emp), a \emph{customer} (\cust),
or an \emph{office} (\room), and each edge represents
either an email \emph{message}  (\msg) or a \emph{visit} (\visit)
from a source node to a target node. Each edge is associated with
the set of timepoints when an interaction occurred.  Such graphs
with static nodes but dynamic edges that are active at multiple
timepoints are commonly used to represent interaction
networks~\cite{DBLP:conf/wsdm/ParanjapeBL17,zhao2010Comm}.

\begin{example}
\label{ex:exist-const}
Assume our temporal graph holds information about
 a publicly traded company. Suppose that employee $v_1$ shared
 confidential information with their colleagues $v_2$ and $v_3$, and that one of them subsequently shared this information with customer $v_4$, potentially constituting insider trading. Assuming that we have no access to the content of the messages, only to when they were sent, can we identify employees who may have leaked confidential information to $v_4$?
 
 Based on graph topology alone, both $v_2$ and $v_3$ could have
  been the source of the information leak to $v_4$.  However, by
  considering the timepoints on the edges, we  observe that there
  is no path from $v_1$ to $v_4$ that goes through $v_3$ and
  visits the nodes in temporal order.  We will represent this
  scenario with the following basic graph pattern (BGP),
  augmented with a temporal constraint:  $$\mathsf{v_1}
  \xrightarrow{y_1} x\xrightarrow{y_2} \mathsf{v_4}: \exists t_1
  \in y_1.T,\, \exists t_2 \in y_2.T:  t_1 \leq t_2$$
 
Here, $\mathsf{v_1}$ and $\mathsf{v_4}$ are node constants, $x$
is a node variable, $y_1$ and $y_2$ are edge variables, and
$y_1.T$ and $y_2.T$ refer to the sets of timepoints associated
with edges $y_1$ and $y_2$.  The temporal
constraint states that there must exist a pair of timepoints
$t_1$, associated with edge $y_1$, and $t_2$, associated with
edge $y_2$, such that $t_1$ occurs before $t_2$.   We refer to
such combinations of BGPs and temporal constraints
as \emph{temporal BGPs}.
\end{example}

The temporal constraint in the above example is
\emph{existential}: it requires the existence of timepoints where
the edges from a BGP matchings are active, so that these
timepoints satisfy some condition (in the example, a simple
inequality).  Existential constraints are typical in the
literature on temporal graph pattern matching
\cite{DBLP:conf/sigmod/GurukarRR15,DBLP:conf/wsdm/ParanjapeBL17,semertzidis2016durable,xu2017time,zfy-temporal-clique,DBLP:conf/edbt/ZufleREF18}.
Various forms of conditions, beyond inequalities on the
timepoints, have been considered.  For example, one may require
that the timepoints belong to a common interval with a given start-
and end-time (``temporal clique'' \cite{zfy-temporal-clique}) or
with a given length (``$\delta$-motifs''
\cite{DBLP:conf/wsdm/ParanjapeBL17}), or one may specify lower
and upper bounds on the gaps between the timepoints
\cite{DBLP:conf/edbt/ZufleREF18}.

In this paper, our goal is to go beyond existential constraints.  Indeed,
many useful temporal constraints are \emph{not} existential.  We give
two examples over temporal graphs such as the one in
Figure~\ref{fig:def2}.
 
\begin{figure*}
    \centering
    \subfloat[BGP Cycle2]{
    \label{fig:bgp-cycle2:bgp}%
        \resizebox{0.35\columnwidth}{!}{
         {\begin{tikzpicture}[->, on grid, node distance=50pt]
            \node[state ] (x) {$x_1:\cust$};
            \node[state, right of=x] (y) {$x_2:\emp$};
            \draw
            (x) edge[bend left] node[above]{$y_1$} (y)
            (y) edge[bend left] node[below]{$y_2$} (x);
         \end{tikzpicture}}
     }
     }
    \quad
      \subfloat[timed automaton $\ta_1$]{
      \label{fig:bgp-cycle2:ta1}
        {\begin{tikzpicture}[->, on grid, node distance=40pt]
                \node[state, initial, accepting] (s0) {$s_0$};
                \node[state, accepting, right of=s0] (s1) {$s_1$};
                \draw
                (s0) edge[bend left] node[above]{$\{y_1\} $} (s1)
                      (s1) edge[loop above] node{$\emptyset$} (s1)
                      (s0) edge[loop above] node{$\emptyset$} (s0)

                (s1) edge[bend left] node[below]{$\{y_2\} $}
                (s0);
        \end{tikzpicture}}
        }
        \qquad
        \subfloat[timed automaton $\ta_2$]{
        \label{fig:bgp-cycle2:ta2}
         {\begin{tikzpicture}[->, on grid, node distance=60pt]
                \node[state, initial, accepting] (s0) {$s_0$};
                \node[state, accepting, right of=s0] (s1) {$s_1$};
                \draw
                (s0) edge[bend left] node[above]{$\{y_1\} \land \xx < 3$} node[below] {$\xx:=0$} (s1)
                     (s1) edge[loop above] node{$\emptyset$} (s1)
                      (s0) edge[loop above] node{$\emptyset$} (s0)
                (s1) edge[bend left] node[below]{$\{y_2\} \land \xx < 3$}
                node[above] {$\xx:=0$}(s0);
        \end{tikzpicture}}
        }
        \subfloat[timed automaton $\ta_3$]{%
            \label{fig:nfa-type-ta3}%
         {\begin{tikzpicture}[->, on grid, node distance=40pt]
                \node[state, initial] (s0) {$s_0$};
                \node[state, right of=s0] (s1) {$s_1$};
                \node[state, accepting, right of=s1] (s2) {$s_2$};
                \draw (s0) edge[loop above] node{$\mathit{\emptyset}$} (s0)
              (s0) edge[above] node{$\{y_1\}$}  (s1)
              (s1) edge[loop above] node{$\neg y_2$} (s1)
              (s1) edge[above] node{$y_2$} (s2)
              (s2) edge[loop above] node{$\mathit{true}$} (s2);
;
        \end{tikzpicture}}
}

  \subfloat[timed automaton $\ta_5$]{
    \label{fig:ta5}%
         \resizebox{0.35\columnwidth}{!}{\begin{tikzpicture}[->, on grid, node distance=40pt]
                \node[state, initial, accepting] (s0) {$s_0$};
                \draw
  (s0) edge[loop above] node{$\emptyset \lor (y_1 \land y_2) $} (s0);
        \end{tikzpicture}}
     }
    \quad
    \subfloat[timed automaton $\ta_6$]{
      \label{fig:ta6}
         \resizebox{0.35\columnwidth}{!}{\begin{tikzpicture}[->, on grid, node distance=45pt]
                \node[state, initial, accepting] (s0) {$s_0$};
                \draw
  (s0) edge[loop above] node{$\emptyset \lor ({y_1} \oplus {y_2} ) $} (s0);
        \end{tikzpicture}}
        }
        \quad
   \subfloat[timed automaton $\ta_7$]{
        \label{fig:ta7}
         \resizebox{0.6\columnwidth}{!}{\begin{tikzpicture}[->, on grid, node distance=50pt]
                \node[state, initial] (s0) {$s_0$};
                \node[state, right of=s0] (s1) {$s_1$};
                \node[state, accepting, right of=s1] (s2) {$s_2$};
                \draw
                (s0) edge[left] node[above]{${y_1} \land {y_2}$} node[below] {$\xx:=0$} (s1)
                (s0) edge[loop above] node{$\mathit{true}$} (s0)
                (s1) edge[loop above] node{${y_1} \land {y_2}$} (s1)
                
                (s1) edge[left] node[above]{$\mathit{true}~\land $} node[below]{$ \xx > 3 $}  (s2)
                
                (s2) edge[loop above] node{$\mathit{true}$} (s2)
                ;
        \end{tikzpicture}}
        }
    \quad
     \subfloat[timed automaton $\ta_8$]{
    \label{fig:ta8}%
         \resizebox{0.33\columnwidth}{!}{\begin{tikzpicture}[->, on grid, node distance=60pt]
        \node[state,accepting, initial]   (s0)                     {$s_0$};
        \draw 
            (s0) edge[loop above] node{$\neg y_1 \lor y_2$} (s0);
        \end{tikzpicture}}
     }
\caption{Example of a temporal BGP and example timed automata
explained in Example~\ref{ex:ta}.}
        \label{fig:bgp-star2}
\end{figure*}
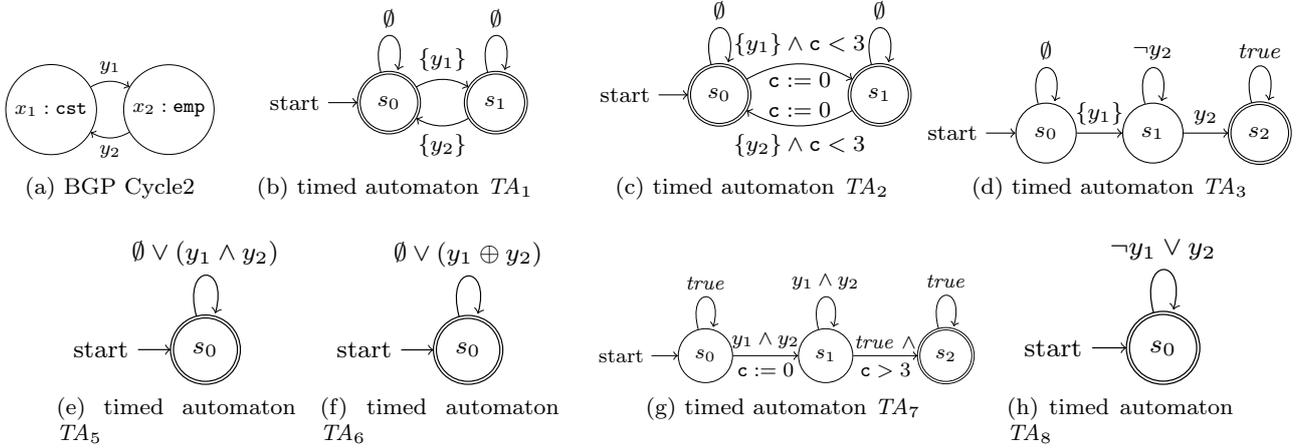

\begin{example}
\label{ex:complex-const1}

When monitoring communication patterns, we may want to look for
  extended interactions between customers and employees.
  Specifically, we are looking for matchings of the BGP shown in
  Figure~\ref{fig:bgp-cycle2:bgp}, where edge variables $y_1$ and
  $y_2$ represent email messages exchanged by customer $x_1$ and
  employee $x_2$.  We impose the temporal constraint that
  $y_1$ and $y_2$ are active in an interleaved, alternating
  fashion: first $y_1$ was active, then $y_2$, then $y_1$ again,
  etc.  This constraint is not existential.  In Figure~\ref{fig:def2}, 
  it is satisfied in
  the communication between $v_5$ and $v_1$, but not between
  $v_7$ and $v_1$.

\end{example}

\begin{example}
\label{ex:complex-contact}

In contact tracing, we may want to look for pairs of employees
  who have shared an office for a contiguous period of time with
  some minimal duration. We are looking for
  matchings of the BGP $x_2:\emp \xrightarrow {y_1:\visit}
  x_1:\room \xleftarrow{y_2:\visit} x_3:\emp $.  As a temporal
  constraint, we impose that there exists a contiguous sequence
  of timepoints in the graph's temporal domain, of duration at
  least, say, 3 time units, in which $y_1$ and $y_2$ were both
  active.  This constraint is, again, not existential.  In
  Figure~\ref{fig:def2}, the only matching of the BGP (involving
  employees $v_2$ and $v_4$ and office $v_8$) does not satisfy
  the constraint; as a matter of fact, $v_2$ and $v_4$ were 
  never active (\ie at the office) at the same time!  Indeed, ``$y_1$
  and $y_2$ are never active at the same time'' would be another
  natural example of a temporal constraint that is not existential.

\end{example} 

In order to express temporal constraints (existential or not), we
need a language.  When the goal is the expression of possibly
complex constraints, 
an obvious approach would
be to use SQL\@.  Indeed, any temporal graph can be
naturally represented by three relations $\Node(\vid,\lab)$,
$\Edge(\eid,\vid_1,\vid_2,\lab)$, and $\Xitab(\eid,\tim)$, where
$\vid$ and $\eid$ are node (vertex) and edge identifiers.
The problem with this approach is that temporal constraints do
not fit well in the SQL idiom.  SQL is certainly expressive
enough, but the resulting expressions tend to be complicated and
hard to optimize.  The alternating communication pattern from
Example~\ref{ex:complex-const1} would be expressed in SQL as
follows:
\label{fig:query-ta}
\begin{lstlisting}
QTA1: WITH matching AS
(SELECT E1.eid AS y1, E2.eid AS y2
 FROM edge E1, edge AS E2
 WHERE E1.dst = E2.src and E2.dst = E1.src),
Succ AS
(SELECT y1, y2, A1.eid AS e1, A2.eid AS e2
 FROM matching, active A1, active A2
 WHERE (A1.eid = y1 OR A1.eid = y2) 
   AND (A2.eid = y1 OR A2.eid = y2)
   AND A1.time<A2.time AND NOT EXISTS
         (SELECT * FROM active A3
          WHERE (A3.eid = y1 or A3.eid = y2)
            AND A1.time<A3.time 
            AND A3.time<A2.time))
SELECT * FROM matching M
WHERE NOT ( EXISTS (SELECT * FROM active WHERE eid = y2)
            AND NOT EXISTS (SELECT * FROM active WHERE eid = y1) )
  AND ( NOT EXISTS (SELECT * FROM active A1, active A2
                    WHERE A1.eid = y1 AND A2.eid = y2)
        OR (SELECT MIN(time) FROM active WHERE eid = y1) <
           (SELECT MIN(time) FROM active WHERE eid = y2) )
  AND NOT EXISTS (SELECT * FROM Succ
                  WHERE M.y1 = y1 AND M.y2 = y2 AND e1 = e2)
\end{lstlisting}
Likewise, for the contiguous-duration office sharing pattern from
Example~\ref{ex:complex-contact}:
\begin{lstlisting}
QTA7: WITH matching AS
(SELECT E1.eid AS y1, E2.eid AS y2
 FROM edge E1, edge AS E2 WHERE E1.dst = E2.dst)
SELECT DISTINCT y1, y2
FROM matching, active A1, active A2, active B1, active B2
WHERE A1.eid = y1 AND A2.eid = y2 AND A1.time = A2.time
  AND B1.eid = y1 AND B2.eid = y2 AND B1.time = B2.time
  AND B1.time - A1.time > 3
  AND NOT EXISTS
      (SELECT * FROM active C
       WHERE A1.time < C.time AND C.time < B1.time
         AND NOT EXISTS (SELECT * FROM active C1, active C2
                         WHERE C1.time = C.time AND C2.time = C.time
                         AND C1.eid = y1 AND C2.eid = y2))
\end{lstlisting}

The hypothesis put forward in this paper is that specification
formalisms used in fields such as complex event recognition
\cite{stijn-cer} or verification of real-time systems
\cite{timed-automata-survey} may be much more suitable for the
expression of complex temporal constraints. In this paper, we
specifically investigate the use of \emph{timed automata}
\cite{timed-automata,timed-automata-survey}.

\begin{example} \label{ex:ta}

Figure~\ref{fig:bgp-star2} shows various examples of timed
  automata that can be applied to matchings of a BGP with two
  edge variables $y_1$ and $y_2$, such as the BGPs considered in
  Examples \ref{ex:complex-const1} and \ref{ex:complex-contact}.
One can think of the automaton as running over the snapshots of
  the temporal graph.  A matching is accepted if there is a run
  such that, after seeing the last snapshot, the automaton is in an accepting state. The 
  edge variables serve as Boolean conditions on the transitions
  of the automaton.  When the edge matched to $y_1$ ($y_2$) is
  active in the current snapshot, the Boolean variable $y_1$ ($y_2$),
  is true.  We use $\emptyset$ as an abbreviation
  for $\neg y_1 \land \neg y_2$, and $\{y_1\}$ for $y_1 \land
  \neg y_2$ (and similarly $\{y_2\}$).

  The alternation constraint of Example~\ref{ex:complex-const1}
  is expressed by $\ta_1$. $\ta_2$ is similar but additionally
  requires that each message gets a reply within $3$ time units
  (a clock $c$ is used for this purpose). The contiguous-duration
  constraint of Example~\ref{ex:complex-contact} is expressed by
  $\ta_7$, also using a clock.  The constraint ``$y_1$ and $y_2$
  are never active together'' is expressed by $\ta_6$; the
  opposite constraint ``$y_1$ and $y_2$ are always active
  together'' by $\ta_5$.  Likewise, $\ta_8$ expresses that $y_2$
  is active whenever $y_1$ is (in SQL, this constraint would
  correspond to a set containment join
  \cite{mamoulis-set-containment-revisited}).  Finally, $\ta_3$
  expresses that $y_1$ has been active strictly
  before the first time $y_2$ becomes active. Also,
  existential constraints such as the one from
  Example~\ref{ex:exist-const} are readily expressible by timed
  automata (see Section~\ref{sec:automata}).

\end{example}

Timed automata offer not only a good balance between
expressivity and simplicity. A temporal constraint expressed by a
timed automaton can also be processed efficiently, as the
iterative state assignment mechanism allows early acceptance and
early rejection of matchings.  In this paper,
we will introduce three algorithms for the evaluation of temporal
BGPs with timed automata as temporal constraints.
The first is a
baseline algorithm intended for offline processing when the
complete history of graph evolution is available at the time of
execution.  The second is an \od algorithm that supports online
query processing when the temporal graph arrives as a stream.
The third is a \pmatch algorithm that speeds up processing by
sharing computation between multiple matches.

We will present an implementation of these algorithms in a
dataflow framework, and will analyze performance trade-offs
induced by the properties of the temporal BGP and of the
underlying temporal graph.  We will also compare performance with main-memory
SQL systems, and will observe that temporal BGPs with
temporal constraints that are not existential can be impractical when
expressed and processed as SQL queries.

\section{Temporal graphs and temporal graph patterns}
\label{sec:model}

We begin by recalling the standard notions of graph and graph pattern used in graph databases~\cite{DBLP:journals/csur/AnglesABHRV17,wood_survey}. 
Assume some vocabulary $L$ of \emph{labels}. We define:

\begin{definition}[Graph]
\label{def:tg1}
A graph is a tuple $(N,E,\rho,\lambda)$, where:
\begin{itemize}
\item $N$ and $E$ are disjoint sets of \emph{nodes} and \emph{edges}, respectively;
\item $\rho: E \to (N \times N)$ indicates, for each edge, its source and destination nodes; and
\item $\lambda: N \cup E \to L$ assigns a label to every node and edge.
\end{itemize}
\end{definition}

Figure~\ref{fig:def2} gives an example of a graph, with 
$N=\{v_1, \ldots, v_7 \}$, $E=\{e_1, \ldots, e_9 \}$,
$\lambda(v_1) = \lambda(v_2) = \lambda(v_3) = \emp$, $\lambda(v_4) = \ldots = \lambda(v_7) = \cust$, and $\lambda(e_1) = \ldots = \lambda(e_9) = \msg$. In this graph, $\rho(e_5) = (v_5, v_1)$.

Next, recall the conventional notion of basic graph pattern (BGP).

\begin{definition}[Basic graph pattern (BGP)]
A BGP is a tuple $(C,X,Y,\rho,\lambda$),
where:
\begin{itemize}
\item $C$, $X$ and $Y$ are pairwise disjoint finite sets of
\emph{node constants}, \emph{node variables}, and \emph{edge
variables}, respectively;
\item $\rho: Y \to (C \cup X)  \times (C \cup X)$ indicates,
for each  edge variable, its source and destination, which can
be a node constant or a node variable; and
\item $\lambda:X \cup Y \to L$ is a partial function, assigning a label from $L$ to some of the variables.
\end{itemize}
\label{def:bgp}
\end{definition}

The fundamental task related to BGPs is to find all matchings in a graph, defined as follows:

\begin{definition}[Matching] \label{defmatch}
A \emph{partial matching} of a BGP
$(C,X,Y,\allowbreak \rho_P,\allowbreak \lambda_P)$ in a graph
$G=(N,E,\rho,\lambda)$ is a function $\mu:Z \to N \cup E$
satisfying the following conditions:
\begin{itemize}
\item
$Z$, the domain of $\mu$, is a subset of $X \cup Y$.
\item
$\mu(Z \cap X) \subseteq N$ and $\mu(Z \cap Y) \subseteq E$.
\item
Let $y$ be an edge variable in $Z$ and let $\rho_P(y)=(x_1,x_2)$.
Then, for $i=1,2$, if $x_i$ is a node variable, then $x_i \in Z$.
Moreover, $\rho(\mu(y)) = (\mu(x_1),\mu(x_2))$, where we agree
that $\mu(c)=c$ for any node constant $c$.
\item
For every $z\in Z$ for which $\lambda_P(z)$ is defined, we have
$\lambda(\mu(z)) = \lambda_P(z)$.
\end{itemize}
If $Z$ equals $X\cup Y$ then $\mu$ is called a (total) \emph{matching}.
\end{definition}

Consider the BGP in Figure~\ref{fig:bgp-cycle2:bgp}.
Evaluating it over the graph in Figure~\ref{fig:def2} yields 7 partial matchings: $v_1 \xrightarrow{e_6} v_5$,  $v_1 \xrightarrow{e_9} v_7$, $v_2 \xrightarrow{e_3} v_4$, $v_3 \xrightarrow{e_4} v_4$, $v_5 \xrightarrow{e_5} v_1$, $v_6 \xrightarrow{e_7} v_1$, $v_7 \xrightarrow{e_8} v_1$,  and 2 total matchings: $v_5 \xrightarrow{e_5} v_1$ $\xrightarrow{e_6} v_5$ and $v_7 \xrightarrow{e_8} v_1$ $\xrightarrow{e_9} v_7$ as total matchings. 

We now present the notion of a \emph{temporal graph} in which edges are associated with sets of timepoints, while nodes persist over time.  Extending our work to temporal property graphs in which both nodes and edges are associated with temporal information, and where  the properties of nodes and edges can change over time~\cite{DBLP:conf/dbpl/MoffittS17}, is an interesting direction for further research. We assume that \emph{timepoints} are strictly positive real numbers and define:
\begin{definition}[Temporal graph]
\label{def:tg}
A temporal graph is a pair $(G,\xi)$, where $G$ is a graph and
$\xi$ assigns a finite set of timepoints to each edge of $G$.
When $e$ is an edge and $t \in \xi(e)$, we say that $e$ is \emph{active} at time $t$.
\end{definition}

In the temporal graph in Figure~\ref{fig:def2}, 
$\rho(e_5) = (v_5, v_1)$ and $\xi(e_5)= \{1,3,5\}$, 
indicating that $v_5$ messaged $v_1$ at the listed timepoints.

To extend the notion of matching to temporal graphs, we enrich BGPs with temporal constraints, defined as follows.

\begin{definition}[Temporal variables, assignments, and constraints]\label{defassign}
Let $V$ be a set of temporal variables. A \emph{temporal
assignment} $\alpha$  on $V$ is a function that assigns a finite
set of timepoints to every variable in $V$. A \emph{temporal
constraint} over $V$ is a set of temporal assignments on $V$.
This set is typically infinite.  When a temporal assignment $\alpha$ belongs to a temporal constraint $\Gamma$, we also say that $\alpha$ \emph{satisfies} $\Gamma$. 
\end{definition}

For the moment, this is a purely semantic definition of temporal constraints; in Section~\ref{sec:automata} we will present how such constraints may be specified using timed automata.



If we have a matching $\mu$ from a BGP in a graph $G$, and we
consider a temporal graph $(G,\xi)$ based on $G$, we automatically obtain a temporal assignment on the edge variables of the BGP\@.  Indeed, each edge variable is matched to an edge in $G$, and we take the set of timepoints of that edge.  Thus, edge variables serve as temporal variables, and we arrive at the following definition:

\begin{definition}[Temporal BGP, matching]
A temporal BGP is a pair $(P,\Gamma)$ where $P$ is a BGP and $\Gamma$ is a temporal constraint over $Y$ (the edge variables of $P$).

Let $(G,\xi)$ be a temporal graph.
Given a matching $\mu$ of $P$ in $G$, we can consider the temporal assignment
$\alpha_\mu$ on $Y$ defined by $$ \alpha_\mu(y) := \xi(\mu(y)) \qquad \text{for  $y \in Y$.} $$

Now a \emph{matching} of the temporal BGP $(P,\Gamma)$ in the temporal graph $(G,\xi)$ is any matching $\mu$ of $P$ in $G$ such that $\alpha_\mu$ satisfies $\Gamma$.
\end{definition}

In the next section, we describe how timed automata such as that in Figure~\ref{fig:bgp-cycle2:ta1} can be used to represent and enforce such constraints.

\section{Expressing Temporal Constraints}
\label{sec:automata}

Our conception of a temporal BGP, as a standard BGP $P$ equipped with a temporal constraint $\Gamma$ on the edge variables of $P$, leaves open how $\Gamma$ is specified. We pursue the idea to use \emph{timed automata}, an established formalism for expressing temporal constraints in the area of verification \cite{timed-automata,timed-automata-survey}.  Timed automata are often interpreted over infinite words, but here we will use them on finite words.

\paragraph{Timed automata} A timed automaton over a finite set $Y$ of variables is an extension of the standard notion of non-deterministic finite automata (NFA), over the alphabet $\Sigma = 2^Y$ (the set of subsets of $Y$).  Recall that an NFA specifies a finite set of states: an initial state, a set of final states, and a set of transitions of the form $(s_1,\theta,s_2)$, where $s_1$ and $s_2$ are states and $\theta$ is a Boolean formula over $Y$.  The automaton reads a word over $\Sigma$, starting in the initial state.  Whenever the automaton is in a current state $s_1$, the next letter to be read is $a$, and there exists a transition $(s_1,\theta,s_2)$ such that $a$ satisfies $\theta$, the automaton can change state to $s_2$ and move to the next letter.  If, after reading the last letter, the automaton is in a final state, the run accepts. If  there is no suitable transition at some point, or if the last state is not final, then the run fails.  A word is accepted if there exists an accepting run.  

The extra feature added by timed automata to the standard NFA apparatus is a finite set $C$ of \emph{clocks}, which can be used to measure time gaps between successive letters in a \emph{timed} word (to be defined momentarily).  Transitions are of the extended form $$ (s_1,\theta,\delta,R,s_2), \eqno (*) $$ where $s_1$, $\theta$ and $s_2$ are as in NFAs; $\delta$ is a Boolean combination of \emph{clock conditions}; and $R$ is a subset of $C$.  Here, by a clock condition, we mean a condition of the form $\xx \leq g$ or $\xx \geq g$, where $\xx$ is a clock and $g$ is a real number constant representing a time gap.

As just mentioned, a timed automaton works over timed words.  A timed word over an alphabet $\Sigma$ is a sequence of the form $$ (t_1,a_1)\dots (t_n,a_n), $$ where each $a_i \in \Sigma$, and $t_1 < \cdots < t_n$ are timepoints. When the automaton is started on the timed word, all clocks are initially set to $t_0 := 0$. For $i=1,\dots,n$ the automaton runs as follows.  Upon reading position $(t_i,a_i)$, every clock has increased by $t_i-t_{i+1}$. Now the automaton can take a transition $(*)$ as above on condition that the current state is $s_1$ and $a_i$ satisfies $\theta$, as before; and, moreover, the current valuation of the clocks satisfies $\delta$.  If this is so, the automaton can change state to $s_2$, move to the next position in the timed word, and must reset to zero all clocks in $R$.  As with NFAs, a run is accepting if it ends in a final state, and a timed word is accepted if there exists an accepting run.

\paragraph{Using timed automata to express temporal BGP constraints} A timed automaton defines the set of timed words that it accepts. But how does it define, as announced, a temporal constraint over $Y$, which is not a set of timed words, but a set of temporal assignments?  This is simple once we realize that a temporal assignment over $Y$, in the context of a temporal graph $H = (G,\xi)$, is nothing but a timed word over $\Sigma = 2^Y$. We can see this as follows. Let $T = \bigcup \{\xi (e) \mid e \in E\}$ be the set of all distinct timepoints used in $H$; we also refer to $T$ as the \emph{temporal domain} of $H$. Let $T=\{t_1,\dots,t_n\}$, ordered as $t_1<\cdots<t_n$. We can then view any temporal assignment $\alpha : Y \to T$ as the timed word $(t_1,a_1)\dots (t_n,a_n)$, where $a_i = \{y \in Y \mid t_i \in \alpha (y)\}$.

\begin{figure}
\caption{$\ta_e$: an existential constraint} 
\label{fig:tae}
    \begin{center}
    \resizebox{0.67\columnwidth}{!}{
    \begin{tikzpicture}[->, on grid, node distance=60pt]
        \node[state, initial] (s0) {$s_0$};
        \node[state, right of=s0] (s1) {$s_1$};
        \node[state, accepting, right of=s1] (s2) {$s_2$};
        \draw (s0) edge[loop above] node{$\mathit{true}$} (s0)
      (s0) edge[above] node{$y_1$} (s1)
      (s1) edge[loop above] node{$\mathit{true}$} (s1)
      (s1) edge[above] node{$y_2$} (s2)
      (s2) edge[loop above] node{$\mathit{true}$} (s2)

      ;
        \end{tikzpicture}
        }
    \end{center}
\end{figure}
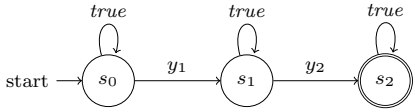

NFAs are a special case of timed automata without any clocks, and this special case is already useful for expressing temporal constraints.
For example, consider the NFA in Figure~\ref{fig:tae} which expresses the existential constraint $ \exists t_1 \in y_1 \exists t_2 \in y_2 : t_1 < t_2$ from Example~\ref{ex:exist-const}. Suppose we instead want to express the existential constraint $ \exists t_1 \in y_1 \exists t_2 \in y_2 : t_1 < t_2-7$. We can do this by introducing a clock $\xx$. When we see $y_1$, we reset the clock ($\xx:=0$).  Then, when we see $y_2$, we check that the clock has progressed beyond the desired $7$ time units ($\xx>7$).
    \begin{center}
     \resizebox{0.80\columnwidth}{!}{
        \begin{tikzpicture}[->, on grid, node distance=70pt]
        \node[state, initial] (s0) {$s_0$};
        \node[state, right of=s0] (s1) {$s_1$};
        \node[state, accepting, right of=s1] (s2) {$s_2$};
        \draw (s0) edge[loop above] node{$\mathit{true}$} (s0)
      (s0) edge[above] node{$y_1$} node[below]{$\xx:=0$} (s1)
      (s1) edge[loop above] node{$\mathit{true}$} (s1)
      (s1) edge[above] node{$y_2 \land \xx>7$} (s2);
        \end{tikzpicture}
        }
    \end{center}
    
Of course, as already argued in the Introduction, timed automata are much more powerful than mere existential constraints. In what follows, we will discuss several algorithms for processing temporal BGPs with temporal constraints given as timed automata.

\section{Algorithms for timed-automaton temporal graph pattern matching}
\label{sec:algorithms}

\subsection{Temporal graph representation }
\label{sec:graph-rep}

Assume that we are given a temporal graph $H=(G,\xi)$ where $G=(N,E,\rho_G,\lambda_G)$. We are also given a temporal BGP $Q=(P,\Gamma)$ where $P=(X,Y,\rho_P,\lambda_P)$ as defined in Section~\ref{sec:model}.
Our goal is to compute all matchings $\mu$ of $Q$ in $H$; recall that this means that $\mu$ must be a matching of $P$ in $G$, and, moreover, the temporal assignment $\alpha_\mu$ must satisfy $\Gamma$ (Definition~\ref{defassign}). We assume that $\Gamma$ is specified as a timed automaton over the alphabet $\Sigma=2^Y$.

We can naturally represent $H$ by the following relations: $\Node(\vid,\lab)$, $\Edge(\eid,\vid_1,\vid_2,\lab)$, and $\Xitab(\eid,\tim)$. 

We recall the natural notion of snapshot from temporal databases:
\begin{definition}[Snapshot] \label{defsnap}
For any timepoint $t_i$ in the temporal domain of $H$, the
\emph{snapshot} of $H$ at time $t_i$ is the subgraph of $G$
induced by all edges that are active at time $t_i$.  We denote
this subgraph by $H_{t_i}$.
\end{definition}
The snapshot can be represented by the
relevant slices of the tables $Node$, $Edge$ and $Active$.



We next present three algorithms for finding the matchings of a temporal BGP in a temporal graph, when the temporal constraint is given by a timed automaton.  We start by presenting our \textbf{baseline} algorithm that operates in two stages. First, it generates all matchings of the BGP; next, it filters out those matchings that violate the temporal constraint.  Our second \textbf{on-demand} algorithm works incrementally. It considers the graph in temporal order, snapshot by snapshot. As time progresses, more edges of the graph are seen so that more and more BGP matchings are found. Also, at each timestep, the possible transitions of the automaton are evaluated to keep track of the possible states for each matching.  For newly found matchings, however, the automaton has to catch up from the beginning.  This catching up is avoided in our third \textbf{partial-match} algorithm, which incrementally maintains all \emph{partial} matches of the BGP, refining them as time progresses.

\subsection{Baseline algorithm}
\label{sec:algorithms:baseline}

Assume a temporal BGP $Q=(P,\Gamma)$, where $\Gamma$ is specified as a timed automaton.  Given a temporal graph $H=(G,\xi)$, we want to find all the matchings of $Q$ in $H$.  We do this in two stages:
\begin{description}
    \item[Find the BGP matchings:] Find all matchings of $P$ in
    $G$ using any of the available algorithms for this task
    \cite{DBLP:journals/sigmod/NgoRR13,amsj_subgraph}.
    \item[Run the automaton:] For each obtained matching $\mu$: 
    \begin{enumerate}[(a)] \item Convert the assignment $\alpha_\mu$ into a timed word over the temporal domain of $H$, as described in Section~\ref{sec:automata}.  We denote this timed word by $\word\mu$.
    \item Check if $\word\mu$ is accepted by automaton $\Gamma$.
    \end{enumerate}
\end{description}

We next describe how the automaton stages (a) and (b) can be done
synchronously, for all matchings $\mu$ in parallel. We use a
table $\States$ that holds triples $(\mu,s,\nu)$, where  $\mu$ is a
matching; $s$ is a state of the automaton; and $\nu$ is 
an assignment of timepoints to the clocks of $\Gamma$.
Since $\Gamma$ is nondeterministic, the same $\mu$ may be
paired with different $s$ and $\nu$.  Naturally, in the initial
content of $\States$, each $\mu$ is paired with state 
$s_0$ and 
$\nu_0$ that maps every clock to 0. 

Let $T=\{t_1,\dots,t_n\}$ with $t_1 < \cdots < t_n$ be the
temporal domain of $H$ as described in Section~\ref{sec:automata}, and
let $t_0 := 0$.  Recall that the active timepoints for each edge
 are stored in the table $\Xitab(\eid,\tim)$.
We obtain $T$ by first sorting $\Xitab$
on time and then scanning through it.  Now during this scan, 
for $i=1,\dots,n$, we do the following:
\begin{enumerate}
\item
Update each $(\mu,s,\nu)$ in $\States$ by
increasing every clock value by $t_i - t_{i-1}$.
\item \label{booleftouterjoin}
Let $Y$, the set of edge variables of $P$, be
$\{y_1,\dots,y_k\}$.
Extend each $(\mu,s,\nu)$ in $\States$ with Boolean values
$b_1,\dots,b_k$ defined as follows: $b_j$ is true if edge $\mu(y_j)$
is active at the current time $t_i$, and false otherwise.
Observe that the bit vector $b_1\dots b_k$ represents the $i$-th
letter of the timed word $\word\mu$.
\item
Join all records $(\mu,s,\nu,b_1\dots b_k)$ from $\States$ with
all transitions $(s_1,\theta,\delta,R,s_2)$ from
$\Gamma$, where the following
conditions are satisfied: $s=s_1$; $b_1\dots b_k$ satisfies
$\theta$; and $\nu$ satisfies $\delta$.
\item
Project every joined tuple on $(\mu,s_2,\nu')$, where $\nu'$ is
$\nu$ but with every clock from $R$ reset to 0.  The resulting
projection is the new content of $\States$. 
\end{enumerate}

\paragraph{Complexity} Each of the above steps can be accomplished by
relational-algebra-like dataflow operations over the $\States$
table.  In particular, step~\ref{booleftouterjoin} is done by
successive left outer joins.  For $j=1,\dots,k$, let $A_j$ be the
$\Xitab$ table, filtered on $\tim=t_i$, and renaming $\eid$ to
$b_j$.  We left-outer join $\States$ with $A_j$ on condition
$y_j=b_j$.  If, in the result, $b_j$ is null, it is
replaced by false; otherwise it is replaced by true. The entire second stage, for a fixed timed automaton, can be implemented in time $O(A + nM)$, where $A$ is the size of the $\it Active$ table, $n$ is the size of the temporal domain, and $M$ is the number of matchings returned from the first stage.

\paragraph{Early acceptance or rejection}
After the iteration for $i=n$, the matchings that are accepted by
the automaton $\Gamma$ are those that are paired in $\States$
with an accepting state.  We may also be able to \textbf{accept results
early:} when, during the iteration, a matching $\mu$ is paired with an
accepting state $s$, and all states reachable from $s$ in the
automaton are also accepting, then $\mu$ can already be output.
On the other hand, when all states reachable from $s$ are
\emph{not} accepting, we can \textbf{reject $\mu$ early.}

\begin{example}
\label{ex:base-line-alg}
Consider again the temporal graph in Figure~\ref{fig:def2}, and suppose that we want to find all cycles of length 2 shown in Figure~\ref{fig:bgp-cycle2:bgp}, under the temporal constraint $TA_2$ shown in Figure~\ref{fig:bgp-cycle2:ta2}. 

  
The first stage of the baseline algorithm identified two matchings,  $\mu=(e_5,e_6)$ and $\mu=(e_8,e_9)$.  These are considered by the timed automaton in the second stage.  

Figure~\ref{fig:state-rel0} shows the $\States$ relation with the
timed words $w_\mu$ at times between 0 to 3. At $t=0$, both
matchings are at $s=s_0$, no clocks have been set, and, since
neither of the matchings has any edges, $b_1=0$ and $b_2=0$. At
time $t=1$, $e_5$ is active, hence the bit $b_1$ for the matching
$(e_5,e_6)$ is set to 1, and, since there is a rule in the timed
automaton, state is updated to $s_1$ and clock $x$ is set to $1$.
Matching $(e_8,e_9)$ does not exist at time $1$, and so no change
is made in that row of the $State$ table. At times $t=1.1$ and
$t=1.9$, neither of the matchings' edges are active, hence the only
change is that the clock is updated for $(e_5,e_6)$. At time
$t=2$, for matching $(e_5,e_6)$, the edge $e_6$ is active and the
clock $x$ is less than $2$, hence we move back to state $s_0$.
For the matching $(e_8,e_9)$, $e_8$ is active so we move to $s_1$
and set the clock. At time $t=3$, $(e_5,e_6)$ continues to
alternate, but for $(e_8,e_9)$ we see that $e_8$ is active, hence,
we set $b_1=1$ and $b_2=0$, and, seeing that the timed automaton
does not have a transition, we drop this matching (shown as
grayed out in Figure~\ref{fig:state-rel0}).  Between times 3--6,
the matching $(e_5,e_6)$ continues alternating between $s_0$ and
$s_1$. From time 7--9, we observe neither of $e_5$ or $e_6$, and
hence no change happens to the state of this matching. The final
output of this algorithm is that matching $(e_5,e_6)$ is accepted
at state $s_0$. 

\begin{figure*}[t!]
    \small
    \centering
\begin{tabular}{r|l|l|l|l|r|l|l|l|l|} 
\cline{2-5}\cline{7-10}
$\textbf{t}$ &\w{$\mu$} &\w{$s$} &\w{$v$} &\w{$b_1b_2$} & \textbf{t} &\w{$\mu$} &\w{$s$} &\w{$v$} &\w{$b_1b_2$}  \\ 
\cline{2-5}\cline{7-10}
\multirow{2}{*}{\textbf{0}} &$e_5,e_6$ &$s_0$ & {[}] & 00 & \multirow{2}{*}{\textbf{1.9}} &$e_5,e_6$ &$s_1$~ & {[}x=.9] & 00 \\ 
\cline{2-5}\cline{7-10}
 &$e_8,e_9$ &$s_0$ & {[}] & 00 &  &$e_8,e_9$ &$s_0$ & {[}] & 00 \\ 
\cline{2-5}\cline{7-10}
\multicolumn{1}{l}{} & \multicolumn{1}{l}{} & \multicolumn{1}{l}{} & \multicolumn{1}{l}{} & \multicolumn{1}{l}{} & \multicolumn{1}{l}{} & \multicolumn{1}{l}{} & \multicolumn{1}{l}{} & \multicolumn{1}{l}{} &  \\ 
\cline{2-5}\cline{7-10}
\multirow{2}{*}{1} &$e_5,e_6$ &$s_1$ & {[}x=0] & 10 & \multirow{2}{*}{\textbf{2}} &$e_5,e_6$ &$s_0$ & {[}x=1] & 01 \\ 
\cline{2-5}\cline{7-10}
 &$e_8,e_9$ &$s_0$ & {[}] & 00 &  &$e_8,e_9$ &$s_1$ & {[}x=0] & 10 \\ 
\cline{2-5}\cline{7-10}
\multicolumn{1}{l}{} & \multicolumn{1}{l}{} & \multicolumn{1}{l}{} & \multicolumn{1}{l}{} & \multicolumn{1}{l}{} & \multicolumn{1}{l}{} & \multicolumn{1}{l}{} & \multicolumn{1}{l}{} & \multicolumn{1}{l}{} &  \\ 
\cline{2-5}\cline{7-10}
\multirow{2}{*}{\textbf{1.1}} &$e_5,e_6$ &$s_1$ & {[}x=.1] & 00 & \multirow{2}{*}{\textbf{3}} &$e_5,e_6$ &$s_1$ & {[}x=0] & 10 \\ 
\cline{2-5}\cline{7-10}
 &$e_8,e_9$ &$s_0$ & {[}] & 00 &  & {\cellcolor[HTML]{C0C0C0}}$e_8,e_9$ &{\cellcolor[HTML]{C0C0C0}}  &{\cellcolor[HTML]{C0C0C0}}  &{\cellcolor[HTML]{C0C0C0}} 10 \\
\cline{2-5}\cline{7-10}
\end{tabular}\caption{Content of the $States$ relation at $t=0, \ldots, 3$, illustrating the execution of the timed automaton in Figure~\ref{fig:bgp-cycle2:ta2} by the baseline algorithm of Section~\ref{sec:algorithms:baseline}.} 
    \label{fig:state-rel0}
\end{figure*}
\end{example}

\subsection{On-demand algorithm}
\label{sec:algorithms:ondemand}

A clear disadvantage of the baseline algorithm is that we must
first complete the first stage (BGP matching on the whole
underlying graph $G$) before we can move to the automaton stage. 
This delay may be undesirable and prohibits returning results
early in situations where the temporal graph is streamed over
time.  We next describe our second algorithm, which works
incrementally by processing \emph{snapshots} in chronological order.

Recall Definition~\ref{defsnap} of snapshots.
We also define:
\begin{definition}[History]
The \emph{history} of $H$ until time $t_i$, denoted $\Hist H{t_i}$, is the union
of all snapshots $H_{t_j}$ for $j=1,\dots,i$.  For $t_0:=0$, we define $\Hist H{t_0}$ to be the empty graph.
\end{definition}

The baseline algorithm is now modified as follows.  We no
longer have a first stage.  Snapshots arrive chronologically at
timepoints $t_1,\dots,t_n$; it is not necessary for the algorithm
to know the entire temporal domain $\{t_1,\dots,t_n\}$ in
advance.  For $i=1,\dots,n$:
\begin{enumerate}
\item
We receive as input the next snapshot $H_{t_i}$.
In previous iterations we have already computed
all matchings of $P$ in the preceding history $\Hist H{t_{i-1}}$.
Using this information and the next snapshot, we compute the new
matchings, i.e., the matchings of $P$ in the current history
$\Hist H{t_i}$ that were not yet matchings of $P$ in the
preceding history.  Incremental BGP
matching is a well-researched topic, and any of the available
algorithms can be used here
\cite{DBLP:journals/vldb/YangW03,gupta1993maintaining,fan2013incremental,DBLP:conf/sigmod/KimSHLHCSJ18}.
\item
We use the table $\States$ as in the baseline algorithm.
For each newly discovered matching $\mu$, we must catch up and
run the automaton from the initial state
on the prefix of $w_\mu$ of length $i-1$.  We add to $\States$
all triples $(\mu,s,\nu)$, such that the configuration $(s,\nu)$ is a possible
configuration of the automaton after reading the prefix.
\item
All matchings we already had remain valid; indeed, if $\mu$ is a
matching of $P$ in $\Hist H{t_{i-1}}$ then $\mu$ is also a
matching of $P$ in $\Hist H{t_i}$.  $\States$ is now
updated for the $i$-th letter of the timed words of all
matchings, new and old, as in the baseline algorithm.
\end{enumerate}

We call this the ``on-demand'' algorithm because the
automaton is run from the beginning, on demand, each time new
matchings are found, in order to catch up with table $\States$
holding the possible automaton configurations.

\begin{example}
\label{ex:on-demand-alg}

Figure~\ref{fig:state-rel1} shows the $State$ relation for the on-demand algorithm, for the same BGP and temporal constraint as in Example~\ref{ex:base-line-alg}. The first time a cycle of length 2 exists in the graph in Figure~\ref{fig:def2} is $t=2$, hence there will be no matching in any iteration before $t=2$ and no temporal automaton would run. At time $t=2$ the incremental matching algorithm finds the matching $(e_5,e_6)$ and passes it to the timed automaton that runs it for $t=0$, $t=1$, $t=1.1$ and $t=1.9$ as was described in Example~\ref{ex:base-line-alg}. 
At time $t=9$, edge $e_9$ is received and gives rise to a new matching $(e_8,e_9)$. At that point, the timed automaton is invoked for all $t<9$. The process is similar to what we described in Example~\ref{ex:base-line-alg}, and the on-demand automaton will eliminate this matching at $t=3$ because no rule in the automaton can be satisfied.  The final output is the same as for the baseline algorithm: matching $(e_5,e_6)$ accepted at state $s_0$. Note that using on-demand algorithm, we can process the graph that arrives as a stream. 

\begin{figure}[t!]
    \small
    \centering
    \begin{tabular}{lllll}
\cline{2-5}
        \multicolumn{1}{l|}{} & \multicolumn{1}{l|}{\w{$\mu$}} & \multicolumn{1}{l|}{\w{$s$}} & \multicolumn{1}{l|}{\w{$v$}} & \multicolumn{1}{l|}{\w{$b_1b_2$}} \\ \cline{2-5} 
\multicolumn{1}{l|}{t =2} & \multicolumn{1}{l|}{$(e_5,e_6)$} & \multicolumn{1}{l|}{$s_0$} & \multicolumn{1}{l|}{[x=1]} & \multicolumn{1}{l|}{01} \\ \cline{2-5} 
 &  &  &  &  \\ \cline{2-5} 
\multicolumn{1}{l|}{t =3} & \multicolumn{1}{l|}{$(e_5,e_6)$} & \multicolumn{1}{l|}{$s_1$} & \multicolumn{1}{l|}{[x=0]} & \multicolumn{1}{l|}{10} \\ \cline{2-5} 
$\dots$ &  &  &  &  \\ \cline{2-5} 
\multicolumn{1}{l|}{t=9} & \multicolumn{1}{l|}{$(e_5,e_6)$} & \multicolumn{1}{l|}{$s_0$} & \multicolumn{1}{l|}{[x=0]} & \multicolumn{1}{l|}{00} \\ \cline{2-5} 
\multicolumn{1}{l|}{} & \multicolumn{1}{l|}{\cellcolor[HTML]{C0C0C0}$(e_8,e_9)$} & \multicolumn{1}{l|}{\cellcolor[HTML]{C0C0C0}} & \multicolumn{1}{l|}{\cellcolor[HTML]{C0C0C0}} & \multicolumn{1}{l|}{\cellcolor[HTML]{C0C0C0}01} \\ \cline{2-5} 
\end{tabular}
\caption{Content of the $States$ relation at $t=0, \ldots, 3$, illustrating the execution of the timed automaton in Figure~\ref{fig:bgp-cycle2:ta2} by the on-demand algorithm of Section~\ref{sec:algorithms:ondemand}.}
    \label{fig:state-rel1}
\end{figure}

\end{example}

\subsection{Partial-match algorithm}
\label{sec:algorithms:partial}

A disadvantage of the \od algorithm is the
catching-up of the automaton on newly found matchings.
Interestingly, we can avoid any catching-up and obtain a fully
incremental algorithm, provided we keep not only the total
matchings of $P$ in the current history, but also all
\emph{partial} matchings.

Specifically, we will work with \emph{maximal} partial matchings:
these are partial matchings that cannot be extended to a strictly
larger partial matching on the same graph.  Now,
for any partial matching $\mu$ of $P$ in $G$, we can
define a timed word $w_\mu$, in the same way as for total
matchings.  Formally, $w_\mu = (t_1,a_1) \dots (t_n,a_n)$, where
now $a_i = \{y \in Y \mid \mu$ is defined on $y$ and $t_i \in
\xi(\mu(y))\}$.  The following property now formalizes how
a fully incremental approach is possible: (proof in supplementary materials)
\begin{proposition} \label{propartial}
Let $\mu$ be a maximal partial matching of $P$ in $\Hist H{t_{i-1}}$, and
let $\mu'$ be a partial matching of $P$ in $\Hist H{t_i}$, such
that $\mu \subseteq \mu'$.  Then the timed words $w_\mu$ and
$w_{\mu'}$ have the same prefix of length $i-1$.
\end{proposition}
\begin{proof}
Let $w_\mu = (t_1,a_1),(t_2,a_2) \dots (t_n,a_n)$ and $w_{\mu'} = (t_1,b_1),(t_2,b_2)
\dots (t_n,b_n)$.  We must show that $a_j=b_j$ for
$j=1,\dots,i-1$.  The containment from left to right is
straightforwardly verified. Indeed, take
$y \in a_j$.  Then $\mu(y)$ is defined and $t_j \in \xi(\mu(y))$.
Since $\mu \subseteq \mu'$, also $\mu'(y)=\mu(y)$ is defined and we see
that $y \in b_j$ as desired.

For the containment from right to left, take $y \in b_j$.
Then $\mu'(y)$ is defined and $t_j \in \xi(\mu'(y))$.  For the
sake of contradiction, suppose $\mu(y)$ would not be defined.
Let $\rho_P(y)=(x_1,x_2)$, and
strictly extend $\mu$ to $\mu''$ by mapping $y$ to
$\mu'(y)$; $x_1$ to $\mu'(x_1)$; and $x_2$ to $\mu'(x_2)$.  Since
$\mu'$ is a partial matching of $P$ in $G$, we know that $\mu'(y)$
is an edge in $G$ from node $\mu'(x_1)$ to node $\mu'(x_2)$.
Moreover, since $t_j \in \xi(\mu'(y))$, the edge $\mu'(y)$ is
present in $\Hist H{t_j}$, so certainly also in $\Hist
H{t_{i-1}}$ since $j \leq i-1$.  Thus, $\mu''$ is a partial
matching of $P$ in
$\Hist H{t_{i-1}}$, contradicting the maximality of $\mu$. We conclude
that $\mu(y)$ is defined, and $y \in a_j$. 
\end{proof}

Concretely, the \emph{partial-match} algorithm incrementally maintains,
for $i=1,\dots,n$, all maximal partial matchings $\mu$ of $\Hist
H{t_i}$, along with the possible configurations $(s,\nu)$ of the automaton
after reading the $i$-th prefix the timed word $w_\mu$.
The triples $(\mu,s,\nu)$ are kept in the table $\States$ as
before.  Initially, $\States$ contains just the \emph{single} triple
$(\emptyset,s_0,\nu_0)$, where $\emptyset$ is the empty partial
matching, and $s_0$ (initial automaton state) and $\nu_0$ (every
clock set to 0) are as in the initialization of the baseline algorithm.
For $i=1,\dots,n$, we receive the next snapshot $H_{t_i}$ and
do the following:
\begin{enumerate}
\item
From previous iterations, $\States$ contains all tuples
$(\mu,s,\nu)$, where $\mu$ is a maximal partial matching of $P$ in
$\Hist H{t_{i-1}}$ and $(s,\nu)$ is a possible configuration of the
automaton on the $i-1$-th prefix of $w_\mu$.  Now, using an
incremental query processing algorithm, compute $\Extend$: the set of all pairs
$(\mu,\mu')$ such that $\mu$ appears in $\States$, $\mu'$ extends
$\mu$, and $\mu'$ is a maximal partial matching of $P$ in $\Hist
H{t_i}$.
\item With $\Extend$ computed in the previous iteration,
update
    \begin{multline*}
    \States := \{(\mu',s,\nu)\mid (\mu,s,\nu) \in \States \ \& \
      \\
      (\mu,\mu') \in \Extend \}
    \end{multline*}
by a project--join query.  By Proposition~\ref{propartial},
$\States$ now contains 
all tuples
$(\mu,s,\nu)$, where $\mu$ is a maximal partial matching of $P$ in
$\Hist H{t_i}$ (as opposed to $\Hist H{t_{i-1}}$)
and $(s,\nu)$ is a possible configuration of the
automaton on the $i-1$-th prefix of $w_\mu$.
\item
Exactly as in the baseline and on-demand algorithms, $\States$ is now
updated for the $i$-th letter of the timed words of all
partial matchings.
\end{enumerate}

Note that in step~1 above, it is possible that $\mu'=\mu$, which
happens when the new snapshot does not contain any edges useful
for extending $\mu$, or when $\mu$ is
already a total matching. On the other hand, when $\mu$ can be extended,
there may be many different possible extensions $\mu'$, and
table $\States$ will grow in size.

\begin{example}
\label{ex:partial-alg}  
We now illustrate the partial matching algorithm for the same BGP and temporal constraint as in Examples~\ref{ex:base-line-alg} and~\ref{ex:on-demand-alg}. 
Figure~\ref{fig:state-rel2} shows the $\States$ relation with the
timed words at $t=0,\ldots,3$. (To streamline presentation, we omit 
edges that are not part of any cycle of length 2, but note that
there are 16 such partial matchings in this relation).  At $t=0$,
we only have one partial matching, denoted by $\emptyset$. At
time $t=1$, $e_5$ is active for the first time, and we create two partial
matchings $(e_5,-)$ and $(-,e_5)$. For $(e_5,-)$, $b_1=1$ and,
since the second edge is not set yet, $b_2=0$.  Based on this,
the automaton will update this matching state to $s_1$ and set
the clock to 0. For $(-,e_5)$, we have $b_1=0$ and $b_2=1$, and,
as there is no transition in the automaton for this situation, this
partial matching is dropped early. At
$t=2$, two new edges $e_6$ and $e_8$ are observed, and $(e_6,-),
(-,e_6),(e_8,-), (-,e_8)$ partial matching are added to $\Extend$.
Additionally, $e_6$ can extend $(e_5,-)$, creating the full
matching $(e_5,e_6)$. In this timepoint, as there is no
transition for
$(-,e_6)$ and $(-,e_8)$, they are rejected. At $t=3$, $e_8$ is
active and the partial matching $(e_8,-)$ is rejected. Another observation is that, at $t=3$ we see $e_5$ again, and so we have $b_1=1, b_2=0$. We thus
drop the partial matching $(e_5,-)$, since no edge can extend this matching.  An early rejection such as this can reduce the computation time for the partial matching algorithm. For matching $(e_5,e_6)$, the algorithm continues as
in Example~\ref{ex:base-line-alg}, producing the same result.
\end{example}

\begin{figure}
    \small
    \centering
        \begin{tabular}{lllll}
        \cline{2-5}
        \multicolumn{1}{l|}{} & \multicolumn{1}{l|}{\w{$\mu$}} & \multicolumn{1}{l|}{\w{$s$}} & \multicolumn{1}{l|}{\w{$v$}} & \multicolumn{1}{l|}{\w{$b_1b_2$}} \\ \cline{2-5} 
\multicolumn{1}{l|}{t=0}   & \multicolumn{1}{l|}{$\emptyset$}    & \multicolumn{1}{l|}{$s_0$}& \multicolumn{1}{l|}{[]}& \multicolumn{1}{l|}{00}      \\ \cline{2-5} 
& &&     & \\ \cline{2-5} 
\multicolumn{1}{l|}{t=1}   & \multicolumn{1}{l|}{$\emptyset$}    & \multicolumn{1}{l|}{$s_0$}& \multicolumn{1}{l|}{[]}& \multicolumn{1}{l|}{00}      \\ \cline{2-5} 
\multicolumn{1}{l|}{}       & \multicolumn{1}{l|}{$(e_5, -)$}     & \multicolumn{1}{l|}{$s_1$}& \multicolumn{1}{l|}{[x=1]}     & \multicolumn{1}{l|}{10}      \\ \cline{2-5} 
\multicolumn{1}{l|}{}       & \multicolumn{1}{l|}{\cellcolor[HTML]{C0C0C0}$(-,e_5)$}  & \multicolumn{1}{l|}{\cellcolor[HTML]{C0C0C0}} & \multicolumn{1}{l|}{\cellcolor[HTML]{C0C0C0}[]}    & \multicolumn{1}{l|}{\cellcolor[HTML]{C0C0C0}01} \\ \cline{2-5} 
$\dots$ &  &  &  &  \\ \cline{2-5} 
\multicolumn{1}{l|}{t=2}    & \multicolumn{1}{l|}{$\emptyset$}    & \multicolumn{1}{l|}{$s_0$}& \multicolumn{1}{l|}{[]}& \multicolumn{1}{l|}{00}      \\ \cline{2-5} 
\multicolumn{1}{l|}{}       & \multicolumn{1}{l|}{$(e_5, -)$}     & \multicolumn{1}{l|}{$s_1$}& \multicolumn{1}{l|}{[x=1]}     & \multicolumn{1}{l|}{00}      \\ \cline{2-5} 
\multicolumn{1}{l|}{}       & \multicolumn{1}{l|}{$(e_6, -)$}     & \multicolumn{1}{l|}{$s_1$}& \multicolumn{1}{l|}{[x=0]}     & \multicolumn{1}{l|}{10}      \\ \cline{2-5} 
\multicolumn{1}{l|}{}       & \multicolumn{1}{l|}{\cellcolor[HTML]{C0C0C0}$(-, e_6)$} & \multicolumn{1}{l|}{\cellcolor[HTML]{C0C0C0}} & \multicolumn{1}{l|}{\cellcolor[HTML]{C0C0C0}[]}    & \multicolumn{1}{l|}{\cellcolor[HTML]{C0C0C0}01} \\ \cline{2-5} 
\multicolumn{1}{l|}{}       & \multicolumn{1}{l|}{$(e_8, -)$}     & \multicolumn{1}{l|}{$s_1$}& \multicolumn{1}{l|}{[x=0]}     & \multicolumn{1}{l|}{10}      \\ \cline{2-5} 
\multicolumn{1}{l|}{}       & \multicolumn{1}{l|}{\cellcolor[HTML]{C0C0C0}$(-, e_8)$} & \multicolumn{1}{l|}{\cellcolor[HTML]{C0C0C0}} & \multicolumn{1}{l|}{\cellcolor[HTML]{C0C0C0}[]}    & \multicolumn{1}{l|}{\cellcolor[HTML]{C0C0C0}01} \\ \cline{2-5} 
\multicolumn{1}{l|}{}       & \multicolumn{1}{l|}{$(e_5,e_6)$}    & \multicolumn{1}{l|}{$s_0$}& \multicolumn{1}{l|}{[x=0]}     & \multicolumn{1}{l|}{01}      \\ \cline{2-5} 
&&&     & \\ \cline{2-5} 
\multicolumn{1}{l|}{t=3}    & \multicolumn{1}{l|}{$\emptyset$}    & \multicolumn{1}{l|}{$s_0$}& \multicolumn{1}{l|}{[]}& \multicolumn{1}{l|}{00}      \\ \cline{2-5} 
\multicolumn{1}{l|}{}       & \multicolumn{1}{l|}{\cellcolor[HTML]{C0C0C0}$(e_5, -)$} & \multicolumn{1}{l|}{\cellcolor[HTML]{C0C0C0}} & \multicolumn{1}{l|}{\cellcolor[HTML]{C0C0C0}[x=2]} & \multicolumn{1}{l|}{\cellcolor[HTML]{C0C0C0}10} \\ \cline{2-5} 
\multicolumn{1}{l|}{}       & \multicolumn{1}{l|}{$(e_6,-)$}      & \multicolumn{1}{l|}{$s_1$}& \multicolumn{1}{l|}{[x=1]}     & \multicolumn{1}{l|}{00}      \\ \cline{2-5} 
\multicolumn{1}{l|}{}       & \multicolumn{1}{l|}{\cellcolor[HTML]{C0C0C0}$(e_8,-)$}  & \multicolumn{1}{l|}{\cellcolor[HTML]{C0C0C0}} & \multicolumn{1}{l|}{\cellcolor[HTML]{C0C0C0}[x=1]} & \multicolumn{1}{l|}{\cellcolor[HTML]{C0C0C0}10} \\ \cline{2-5} 
\multicolumn{1}{l|}{}       & \multicolumn{1}{l|}{$(e_5,e_6)$}    & \multicolumn{1}{l|}{$s_1$}& \multicolumn{1}{l|}{[x=1]}     & \multicolumn{1}{l|}{10}      \\ \cline{2-5} 
\end{tabular}
\caption{Content of the $States$ relation at $t=0, \ldots, 3$, illustrating the execution of the timed automaton in Figure~\ref{fig:bgp-cycle2:ta2} by the partial matching algorithm of Section~\ref{sec:algorithms:partial}.}
    \label{fig:state-rel2}
\end{figure}

\subsection{Avoiding quadratic blowup}
\label{sec:par-good}

A well-known problem with partial BGP matching, in the
non-temporal setting, is that the number of partial matchings may
be prohibitively large.

For a simple example, consider matching a path of length 3, $x_1
\xrightarrow{y_1} x_2 \xrightarrow{y_2} x_3 \xrightarrow{y_3}
x_4$, in some graph $G$.  Note that any edge in $G$ gives rise to
a partial matching for $y_1$, $y_2$, and $y_3$.  What is
worse, however, is that any \emph{pair} of edges gives rise to a
partial matching for $y_1$ and $y_3$ together.  We thus
immediately get a quadratic number of partial matchings,
irrespective of the actual topology of the graph $G$.  For
example, $G$ may have no 3-paths at all, or even no 2-paths.
Such a quadratic blowup may not occur for $y_1$ and $y_2$
together.  Indeed, since $y_1$ and $y_2$ form a connected
subpattern, only pairs of \emph{adjacent} edges give rise to a
partial matching.

Of course, in the above example, $G$ may still have many 2-paths
but very few 3-paths, so connectivity is not a panacea.  Still,
we may expect connected subpatterns to have a number of partial
matchings that is more in line with the topology of the graph.
At the very least, working only with connected subpatterns avoids
generating the Cartesian product of sets of partial matchings of
two or more disconnected subpatterns.

Interestingly, in the temporal setting, the very presence of a
temporal constraint (timed automaton) may avoid disconnected
partial matchings.  This happens when the temporal constraint
enforces that only partial matchings of connected subpatterns can
ever satisfy the constraint, allowing early rejection of when partial
matchings of disconnected subpatterns.  We can formalize the above hypothesis as follows. 

Consider a temporal BGP $Q=(P,\Gamma)$.
As usual, let $Y$ be the set of edge variables of $P$.
Consider a total ordering $<$ on $Y$.  We say that:
\begin{description}
\item[$<$ is connected with respect to $P$] if, for every
  $y \in Y$, the subgraph of $P$ induced by all
edge variables $z \leq y$ is connected.
\item[$<$ is compatible with $\Gamma$] if, for any $y_1 < y_2$
in $Y$, and any timed word $w$ satisfying $\Gamma$ in which both $y_1$
and $y_2$ appear, the first position in $w$ where $y_1$ appears
does not come after the first position where $y_2$ appears.
\end{description}

Now, when a connected, compatible ordering is available, we can
modify the partial-match algorithm in the obvious manner so as
\emph{to focus only on partial matchings based on the subsets of variables
$\{y_1,\dots,y_j\}$, for $1 \leq j \leq n$}.  By the connectedness
property, we avoid Cartesian products.  Moreover, by the compatibility
property, we do not lose any outputs.

As a simple example, consider the path of length 3 BGP and the timed automaton $\Gamma$ from Figure~\ref{fig:all-nfa}. The  ordering $y_1 <
y_2 <y_3$ is connected with respect to $P$, and is compatible with
$\Gamma$.  So our theory would predict
the partial matching algorithm to work well for this temporal BGP
$(P,\Gamma)$.  We will show effectiveness of the partial matching algorithm in Section~\ref{sec:experiments:density}.

Whether or not an ordering
of the edge variables is connected with respect to $P$ is
straightforward to check, by a number of graph connectivity
tests.  Moreover, when $\Gamma$ is given by a timed automaton,
it also possible to effectively check whether an ordering is
compatible with $\Gamma$.

\paragraph{Verifying compatibility}  We offer the following algorithm for verifying that an ordering $y_1 < \cdots < y_m$ is compatible with a temporal constraint $\Gamma$, specified by a timed automaton.
\begin{enumerate}
  \item
    Compute an automaton defining the
    intersection of $\Gamma$ with all regular languages of the
    form $$ (\neg y_j \land \neg y_i)^* \cdot (y_j \land \neg y_i)
    \cdot \mathit{true}^* \cdot y_i \cdot \mathit{true}^*, $$
    for $1 \leq i < j \leq n$.
    These languages contain
    the words where both $y_i$ and $y_j$ appear, but $y_j$
    appears first, which we do not want when $i<j$.
  \item
    The resulting timed automaton should represent the empty
    language, i.e., should not accept any timed word.
\end{enumerate}
Effective algorithms for computing the intersection of timed
automaton and for emptiness checking are known
\cite{timed-automata}.  Note that it actually suffices here to
intersect a timed automaton ($\Gamma$) with an NFA (the union of
the regular languages from step~1).
An interesting question for further research is to
determine the precise complexity of the following
problem:
\begin{itemize}[\noindent]
  \item[{\bf Problem:}] Compatible and connected ordering
  \item[{\bf Input:}] A temporal BGP $(P,\Gamma)$
  \item[{\bf Output:}] An ordering of the edge variables that is
    connected w.r.t.\ $P$ and compatible with $\Gamma$, or `NO'
    if no such ordering exists.
\end{itemize}

\section{Implementation}
\label{sec:impl}

The algorithms described in Section~\ref{sec:algorithms} have
been implemented using Rust and the \texttt{Itertools}
library~\cite{rust-itertools} as a single-threaded application.
Our algorithms are easy to implement using any system supporting the dataflow model such as Apache Spark~\cite{DBLP:journals/cacm/ZahariaXWDADMRV16}, Apache Flink~\cite{carbone2015apache}, Timely Dataflow \cite{murray2013naiad}, or Differential Dataflow~\cite{mcsherry2013differential}. 

A temporal graph is stored on disk as relational data in CSV
files $\it Node$, $\it Edge$, and $\it Active$. In the initial stage of the program, we load all data into memory, loading edges into two hash-tables with \insql{vid1} and \insql{vid2} as keys. 
We added a ``first'' meta-property field to the $Edge$ relation and use it for lazy evaluation of matchings in the baseline and on-demand algorithms.  


\paragraph{BGP matching} We implement BGP matching as a select-project-join query. For
cyclic BGPs such as triangles and rectangles, we use worst-case
optimal join~\cite{amsj_subgraph}, meaning that instead of
the traditional pairwise join over the edges, we use a
vertex-growing plan.  We use a state-of-the art method in our
implementation but note that (nontemporal) BGP matching in itself is not the focus of this paper, and so any other BGP matching algorithm  can be used in conjunction with the timed automata-based methods we describe. 

On-demand and partial matching algorithms are both designed to work in online mode, computing new matchings at each iteration. To implement online matching  for the on-demand algorithm, we build on join processing in streams~\cite{xie2007survey}. 
We can use information from the temporal constraint to
avoid useless joins in the incremental computation of
matchings.  For example, consider two edge variables $y$ and
$z$ coming from the BGP\@.  With $E$ the current history of
active edges and $\Delta E$ the edges from the new snapshot, we
must in principle update the join of $\rho_y(E)$ with $\rho_z(E)$ by
three additional joins
$\rho_y(E) \Join \rho_z(\Delta E)$;
$\rho_y(\Delta E) \Join \rho_z(E)$; and
$\rho_y(\Delta E) \Join \rho_z(\Delta E)$.  When the
temporal constraint implies, for example,
that $y$ is never active before $z$, the first of these three
additional joins can be omitted.  Such order information can be
inferred from a timed automaton using similar techniques already
described in the paragraph on verifying compatibility
in Section~\ref{sec:par-good}.


\paragraph{Timed automata} We represent a timed automaton as a relation
$\mathit{Automaton}(s_c, \theta, \delta, R,\allowbreak s_n)$,
in which each tuple corresponds to a transition from the current state $s_c$ to the next state $s_n$.  For example, the timed automaton of Figure~\ref{fig:bgp-cycle2:ta2} is represented as follows:

\begin{center}
\small 
\begin{tabular}{|c|c|c|c|c|}
\hline
\w{$s_c$} & \w{$\theta$} & \w{$\delta$} & \w{$R$} & \w{$s_n$} \\ \hline
$0$ & $00$ & $true$ & [] & $0$ \\ \hline
$0$ & $10$ & $c.0 < 3$ & [0] & $1$ \\ \hline
$1$ & $01$ & $c.0 < 3$ & [0] & $0$ \\ \hline
$1$ & $00$ & $true$ & [] & $1$ \\ \hline
\end{tabular}
\end{center}

The specification of the timed automaton is 
loaded into memory as a hash table, with ($s_c$, $\theta$) as the key. 
The timed word $\theta$ (see Section~\ref{sec:automata}), is
encoded as a bitset. For example, in the timed automaton in
Figure~\ref{fig:bgp-cycle2:ta2}, we encode $y_1 \land \neg y_2$
as $10$, where the first bit corresponds to $y_1$ and the second
to $y_2$. If the transition condition is $true$ then, for a
matching with two variables, we add 4 rows to $\it Automaton$, one for each $00$, $01$, $10$, and $11$.
Using bitsets makes automaton transitions efficient, as we will show in Section~\ref{sec:experiments:automata}. 
Table $\it Automaton$ also stores the clock acceptance condition $\delta$, and a nested field $R$ with an array of clocks to be reset during the transition to the next state.  
To update the state of a matching, we execute a hash-join
followed by a projection between $\it Automaton$ and $\it States$.



Updating the clock for each matching will be computationally expensive. Instead, during the automaton transition, for each matching,  we store the current time (of last snapshot visited) value for that clock instead of setting it to zero. This way, instead of updating all clocks in every iteration, we can just get the correct value of the clock when needed and compute the current value of the clock by subtracting the value of the clock from the current time. 

In many temporal graphs, due to the nature of their evolution, most edges appear for the first time during the last few snapshots.
To optimize performance we implemented a simple but effective
optimization for our
baseline  and  on-demand algorithms: when the initial state of the
timed automaton self-loops on the empty letter, we
will not
run on a matching until at least one of its edges is seen. This
can be determined using the ``first'' meta-property of the $\it Edge$ relation.  
This optimization is not necessary for the partial matching
algorithm, where it is essentially already built-in.

We also implement the early acceptance and early rejection optimizations.

\begin{figure*}[ht!]
    \centering
    \begin{minipage}{0.45\textwidth}
        \centering
    \resizebox{0.9\columnwidth}{!}{
         \begin{tikzpicture}[->, on grid, node distance = 70pt]
            \node[state, initial, accepting] (s0) {$s_0$};
            \node[state, right of = s0,accepting] (s1) {$s_1$};
            \node[state, right of = s1,accepting] (s2) {$\dots$};
            \node[state, right of = s2,accepting] (s3) {$s_{m-1}$};
            \draw (s0)
             (s0) edge[loop above] node{$\emptyset   $} (s250)
              (s1) edge[loop above] node{$\emptyset $} (s1)
              (s2) edge[loop above] node{$\emptyset$} (s2)
              (s3) edge[loop above] node{$\emptyset$} (s3)
            
              (s0) edge[above] node{$\{y_1\}$} (s1)
              (s1) edge[above] node{$\dots$} (s2)
              (s2) edge[above] node{$\{y_{m-1}\}$} (s3)
              (s3) edge[below, bend left = 20] node{$\{y_m\} $} (s0);
            \end{tikzpicture}}
    \caption{$\ta_0$ generalizes $\ta_1$, specifying that edges
    should appear repeatedly in the given temporal order.}
    \label{fig:inc-nfa}%
    \end{minipage}\hfill
    \begin{minipage}{0.45\textwidth}
        \centering
          \resizebox{0.9\columnwidth}{!}{
        \begin{tikzpicture}[->, on grid, node distance = 60pt]
    \node[state, initial] (s0) {$s_0$};
    \node[state, right of = s0] (s1) {$s_1$};
    \node[state, right of = s1] (s2) {$s_2$};
    \node[state, right of = s2,accepting] (s3) {$s_3$};
    \draw (s0)
  (s0) edge[loop above] node{$\emptyset$} (s0)
  (s1) edge[loop above] node{$\neg y_2 \land \neg y_3$} (s1)
  (s2) edge[loop above] node{$\neg y_3$} (s2)
  (s3) edge[loop above] node{$\mathit{true}$} (s3)

  (s0) edge[above] node{$\{y_1\}$} (s1)
  (s0) edge[below, bend right = 18] node{$y_1 \land y_2  \land \neg y_3$} (s2)
  (s0) edge[below, bend right = 34] node{$y_1 \land y_2 \land y_3$} (s3)
  (s1) edge[below, bend right= 18] node{$y_2 \land y_3$} (s3)
  (s1) edge[above] node{$y_2 \land \neg y_3$} (s2)
  (s2) edge[above] node{$y_3$} (s3);
    \end{tikzpicture}
    }        \caption{$\ta_4$: Each of $y_1$, $y_2$ and $y_3$ is active at some point, with first time $y_1$ $\leq$ first time $y_2$ $\leq$ first time $y_3$.}
      \label{fig:all-nfa}
    \end{minipage}
\end{figure*}

\section{Experiments}
\label{sec:experiments}

We now describe an extensive experimental evaluation with several real datasets and temporal BGPs, and demonstrate that using timed automata is practical.  We investigate the relative performance of our methods, and compare them against two state-of-the-art in-memory relational systems, \duck~\cite{DBLP:conf/sigmod/RaasveldtM19} and \hyper~\cite{DBLP:conf/sigmod/0001MK15,DBLP:journals/pvldb/Neumann11}.


{\bf In summary,} we observe that the \od and \pmatch algorithms are effective at reducing the running time compared to the baseline. Interestingly, while no single algorithm performs best in all cases, the trade-off in performance can be explained by the properties of the dataset, of the BGP, and of the temporal constraint.  Our results indicate that \pmatch  is most efficient for acylic BGPs such as paths of bounded length, while \od performs best for cyclic BGPs such as triangles, particularly when evaluated over sparse graphs.  Interestingly, the performance gap between \od and \pmatch is reduced with increasing graph density or BGP size, and \pmatch outperforms \od in some cases.
We also show that algorithm performance is independent of timed automaton size and of the number of clocks.

We show that our methods substantially outperform state-of-the-art relational implementations in most cases.  We also demonstrate that temporal BGPs are more concise than the corresponding relational queries, pointing to better usability of our approach.

\paragraph{Experimental setup} Our algorithms were executed as single-thread Rust programs, running on a single cluster node 
with an Intel Xeon Platinum 8268 CPU, using the Slurm scheduler~\cite{yoo2003slurm}.  We used \duck v.0.3.1 and the \hyper API 0.0.14109 provided by Tableau\footnote{https://help.tableau.com/current/api/hyper\_api}. All systems were run with 32GB of memory on a single CPU.   Execution time of \duck and \hyper 
includes 
parsing, optimizing and executing the SQL query, and does not include the time to create database tables and load them into memory. Similarly, execution time of our algorithms includes 
loading the timed automaton and executing the corresponding algorithm. All execution times are averages of 3 runs; the coefficient of variation of the running times was under 10\% in most cases, and at most  12\%. 


\begin{remark} Since we are dealing with graph data and
  BGPs, one may ask why we implemented our
  algorithms in a dataflow environment, and compare to relational
  systems.  Why not work on top of a graph database system, and
  compare to graph databases?  The reason is that a BGP
  is, in essence, a multiway join query, for which the best
  performance is realized with the help of worst-case optimal join algorithms,
  or relational query processors with very good optimization.  It
  is exactly with respect to these two environments that we
  conduct our experiments.  On the other hand, the main advantage of graph
  database systems is their support of reachability queries or
  regular path queries, which are not part of our basic notion of
  BGP\@.  Rather, our contributions lie in expressive 
  temporal filtering of the matchings of a BGP, for which our
  experimental set-up provides a suitable empirical evaluation.
\end{remark}

\paragraph{Datasets.} Experiments were conducted on 4 real datasets, summarized in Table~\ref{tab:data:summary}, where we list the number of distinct nodes and edges, temporal domain size (``snaps''), the number of active edges across snapshots (``active''), structural density (``struct'', number of edges in the graph, divided by number of edges that would be present in a clique over the same number of nodes), and temporal density (``temp'', number of timepoints during which an edge is active, divided by temporal domain size, on average).
\begin{table*}
\begin{center}
\small 
\begin{tabular}{lcccccc}
 & \multicolumn{1}{l}{} & \multicolumn{1}{l}{} & \multicolumn{1}{l}{} & \multicolumn{1}{l}{} & \multicolumn{2}{c}{\textbf{density}} \\ \hline
\multicolumn{1}{|l|}{} & \multicolumn{1}{c|}{\textbf{nodes}} & \multicolumn{1}{c|}{\textbf{edges}} & \multicolumn{1}{c|}{\textbf{active}} & \multicolumn{1}{c|}{\textbf{snaps}} & \multicolumn{1}{c|}{\textbf{struct}} & \multicolumn{1}{c|}{\textbf{temp}} \\ \hline
\multicolumn{1}{|l|}{\textbf{EPL}} & \multicolumn{1}{c|}{50} & \multicolumn{1}{c|}{1500} & \multicolumn{1}{c|}{35K} & \multicolumn{1}{c|}{25} & \multicolumn{1}{c|}{0.6} & \multicolumn{1}{c|}{0.93} \\ \hline
\multicolumn{1}{|l|}{\textbf{Contact}} & \multicolumn{1}{c|}{541} & \multicolumn{1}{c|}{3349} & \multicolumn{1}{c|}{21K} & \multicolumn{1}{c|}{48} & \multicolumn{1}{c|}{0.16} & \multicolumn{1}{c|}{0.13} \\ \hline

\multicolumn{1}{|l|}{\textbf{Email}} & \multicolumn{1}{c|}{776} & \multicolumn{1}{c|}{65K} & \multicolumn{1}{c|}{1.9M} & \multicolumn{1}{c|}{800} & \multicolumn{1}{c|}{0.1} & \multicolumn{1}{c|}{0.03} \\ \hline
\multicolumn{1}{|l|}{\textbf{FB-Wall}} & \multicolumn{1}{c|}{46K} & \multicolumn{1}{c|}{264K} & \multicolumn{1}{c|}{856K} & \multicolumn{1}{c|}{850} & \multicolumn{1}{c|}{0.0001} & \multicolumn{1}{c|}{0.003} \\ \hline
\end{tabular}
\end{center}
\caption{Description of the experimental datasets.}
\label{tab:data:summary}
\end{table*}

{\bf EPL}, based on the English soccer dataset~\cite{curley2016engsoccerdata}, represents 34,800 matches between 50 teams over a 25-year period. We represent this data as a temporal graph with 1-year temporal resolution, where each node corresponds to a team and a directed edge connects a pair of teams that played at least one match during that year.  The direction of the edge is from a team with the higher number of goals to the team with the lower number of goals in the matches they played against each other that year; edges are added in both directions in the case of a tied result.   This is the smallest dataset in our evaluation, but it is very dense both structurally and temporally.

{\bf Contact} is based on trajectory data of individuals at the University of Calgary over a timespan of 4 hours~\cite{ojagh2021person}. We created a bipartite graph with 500 person nodes and 41 location nodes, where the existence of an edge from a person to a location indicates that the person has visited the location. The original dataset records time up to a second.  To make this data more realistic for a contact tracing application, we made the temporal resolution coarser, mapping timestamps to 5-min windows, and associated individuals with locations where they spent at least 2.5 min. 

{\bf Email}, based on a dataset of email communications within a large European research institution~\cite{DBLP:journals/tkdd/LeskovecKF07}, represents about 1.9M email messages exchanged by 776 users over an 800-day period, with about 65K distinct pairs of users exchanging messages.  This dataset has  high structural density (10\% of all possible pairs of users are connected at some point during the graph's history), and intermediate temporal density (3\%).

{\bf FB-Wall}, derived from the Facebook New Orleans user network dataset~\cite{viswanath-2009-activity}, represents wall posts of about 46K users over a 850-day period, with 264K unique pairs of users (author / recipient of post).  This graph has the largest temporal domain in our experiments, and it is sparse, both structurally and temporally. 

We also use synthetic datasets to study the impact of data characteristics on performance, and describe them in the relevant sections.

\subsection{Relative performance of the algorithms}
\label{sec:experiments:constr}

In our first set of experiments, we evaluate the relative performance of the baseline (Section~\ref{sec:algorithms:baseline}), \od (Section~\ref{sec:algorithms:ondemand}), and \pmatch (Section~\ref{sec:algorithms:partial}) algorithms.  Note that the baseline algorithm can only be used when a graph's evolution history is fully available (rather than arriving as a stream), and that \pmatch is only used when the matching is guaranteed to be connected (Section~\ref{sec:par-good}).

We use the BGP that looks for paths of length 2, with timed
automata $\ta_1$, $\ta_2$ and $\ta_3$ from
Figure~\ref{fig:bgp-star2}.
Automaton $\ta_3$ is interesting for 
showing the impact of early acceptance and rejection on
performance.   


\begin{figure}[t!]
\centering
\subfloat[EPL, 47K BGP matches ]{%
\label{fig:nfa-type:epl}%
\includegraphics[width = 0.5\columnwidth]{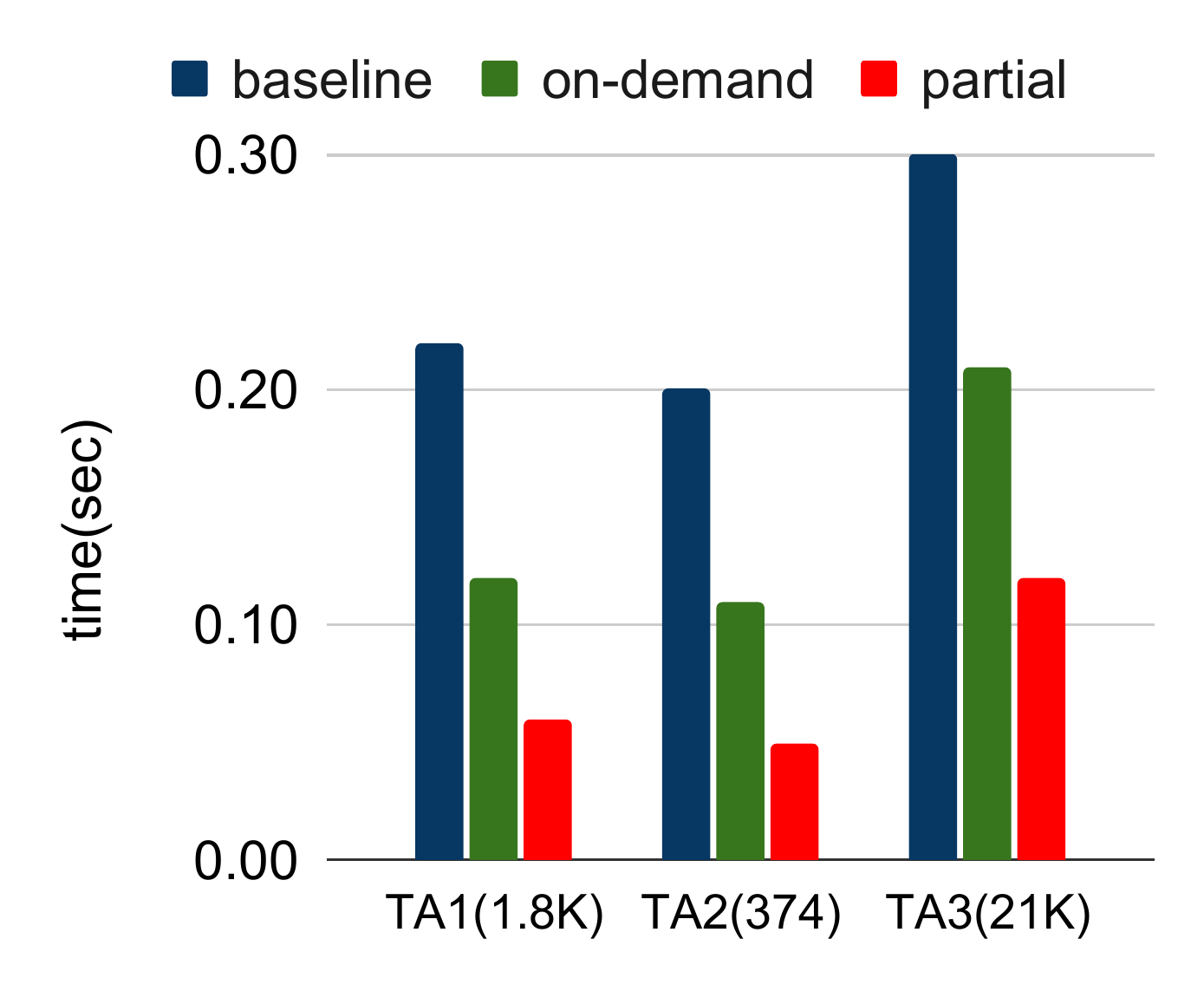}}
\subfloat[Email, 862K BGP matches]{%
\label{fig:nfa-type:euemail}%
\includegraphics[width = 0.5\columnwidth]{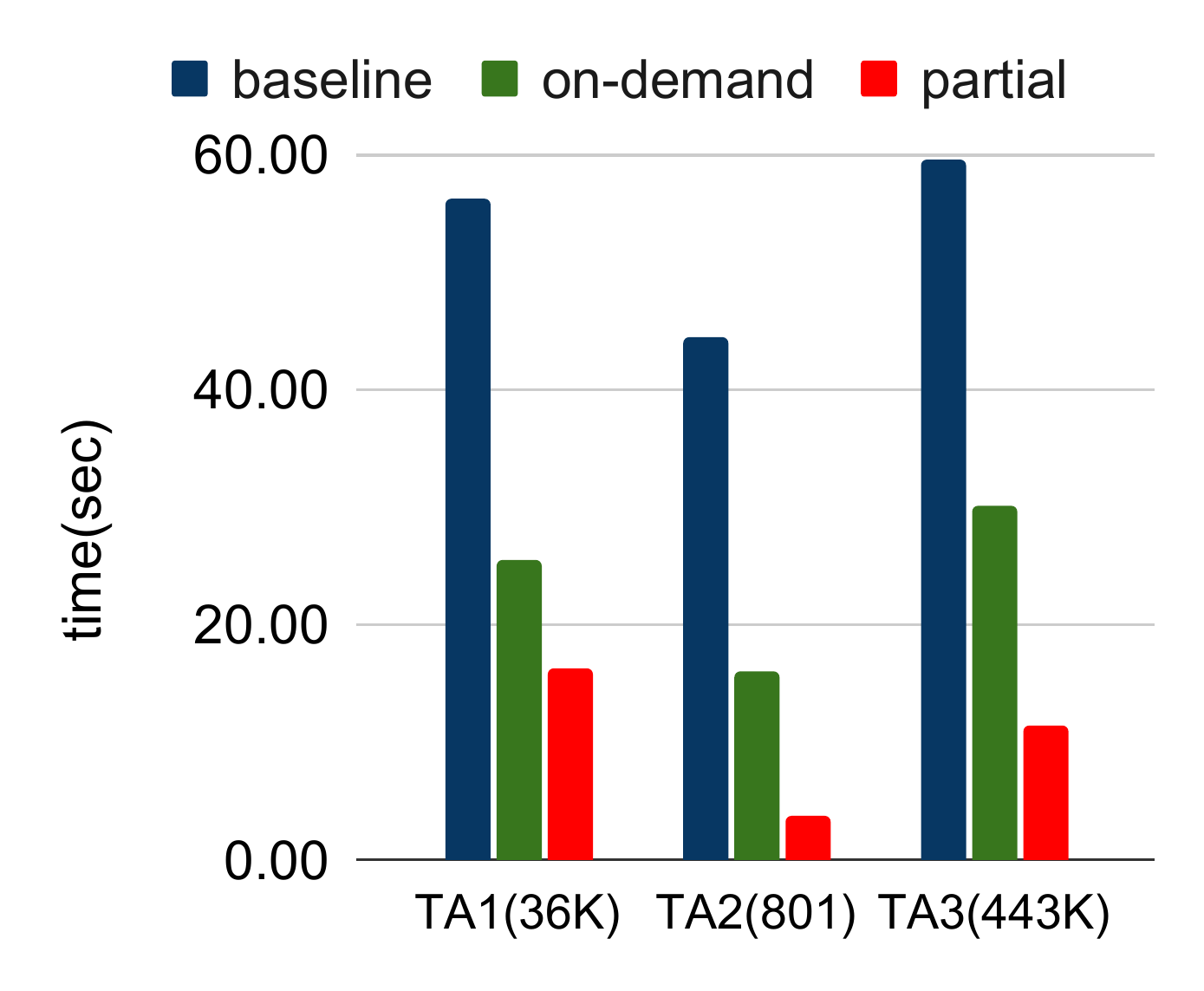}}
\caption{Running time for path of length 2 with timed automata $\ta_1, \ta_2,
\ta_3$.}
\label{fig:nfa-type-time}
\end{figure}

\begin{table*}[h!]
\centering 
\small 
\caption{Relative performance of baseline, \od and \pmatch, for different BGPs and automaton $\ta_0$. }
\label{tab:diff-pattern}
\begin{tabular}{lrrrrrr}

\textbf{EPL} & \multicolumn{2}{c|}{\# of matches} & \multicolumn{4}{c}{time (sec)} \\ \hline
\multicolumn{1}{l|}{pattern} & BGP & \multicolumn{1}{r|}{BGP+TA} & match & baseline & on-demand & partial \\ \hline
\multicolumn{1}{l|}{path2} & 47K & \multicolumn{1}{r|}{1.8K} & 0.01 & 0.22 & 0.12 & \textbf{0.06} \\
\multicolumn{1}{l|}{path3} & 1.5M & \multicolumn{1}{r|}{4.6K} & 0.21 & 5.11 & 3.66 & \textbf{0.32} \\
\multicolumn{1}{l|}{cycle2} & 1.1K & \multicolumn{1}{r|}{35} & 0.01 & \textbf{0.03} & \textbf{0.01} & 0.02 \\
\multicolumn{1}{l|}{cycle3} & 35K & \multicolumn{1}{r|}{66} & 0.02 & 0.12 & \textbf{0.09} & 0.36 \\
\multicolumn{1}{l|}{cycle4} & 1.1M & \multicolumn{1}{r|}{106} & 0.22 & 3.31 & 2.92 & \textbf{0.53} \\
 & \multicolumn{1}{l}{} & \multicolumn{1}{l}{} & \multicolumn{1}{l}{} & \multicolumn{1}{l}{} & \multicolumn{1}{l}{} & \multicolumn{1}{l}{} \\
\textbf{Email} & \multicolumn{2}{c|}{\# of matches} & \multicolumn{4}{c}{time (sec)} \\ \hline
\multicolumn{1}{l|}{pattern} & BGP & \multicolumn{1}{r|}{BGP+TA} & match & baseline & on-demand & partial \\ \hline
\multicolumn{1}{l|}{path2} & 862K & \multicolumn{1}{r|}{36K} & 0.25 & 56.27 & 25.56 & \textbf{16.35} \\
\multicolumn{1}{l|}{path3} & 42M & \multicolumn{1}{r|}{126K} & 41 & 1475.20 & 766.33 & \textbf{74.77} \\
\multicolumn{1}{l|}{cycle2} & 18K & \multicolumn{1}{r|}{843} & 0.12 & 0.92 & \textbf{0.54} & 2.41 \\
\multicolumn{1}{l|}{cycle3} & 205K & \multicolumn{1}{r|}{309} & 0.19 & 5.13 & \textbf{3.49} & 14.34 \\
\multicolumn{1}{l|}{cycle4} & 8.4M & \multicolumn{1}{r|}{1352} & 4.38 & 196.50 & 125.85 & \textbf{92.40} \\
 & \multicolumn{1}{l}{} & \multicolumn{1}{l}{} & \multicolumn{1}{l}{} & \multicolumn{1}{l}{} & \multicolumn{1}{l}{} & \multicolumn{1}{l}{} \\
\textbf{Contact} & \multicolumn{2}{c|}{\# of matches} & \multicolumn{4}{c}{time (sec)} \\ \hline
\multicolumn{1}{l|}{pattern} & BGP & \multicolumn{1}{r|}{BGP+TA} & match & baseline & on-demand & partial \\ \hline
\multicolumn{1}{l|}{I-Star2} & 10.1K & \multicolumn{1}{r|}{0} & 0.01 & 0.03  & 0.03 & \textbf{0.01} \\
\multicolumn{1}{l|}{O-Star2} & 207K & \multicolumn{1}{r|}{0} & 0.03 & 0.84 & 0.79 & 0.67 \\
\multicolumn{1}{l|}{I-Star3} & 21.4K & \multicolumn{1}{r|}{0} & 0.01 & 0.07 & 0.06 & \textbf{0.03} \\
\multicolumn{1}{l|}{O-Star3} & 10.7M & \multicolumn{1}{r|}{0} & 2.86 & 36.24 & 35.26 & \textbf{1.73} \\
 & \multicolumn{1}{l}{} & \multicolumn{1}{l}{} & \multicolumn{1}{l}{} & \multicolumn{1}{l}{} & \multicolumn{1}{l}{} & \multicolumn{1}{l}{} \\
\textbf{FB-wall} & \multicolumn{2}{c|}{\# of matches} & \multicolumn{4}{c}{time (sec)} \\ \hline
\multicolumn{1}{l|}{pattern} & BGP & \multicolumn{1}{r|}{BGP+TA} & match & baseline & on-demand & partial \\ \hline
\multicolumn{1}{l|}{path2} & 4.4M & \multicolumn{1}{r|}{681K} & 1.08 & 839.30 & 387.31 & \textbf{318.14} \\
 \multicolumn{1}{l|}{path3} & 91M & \multicolumn{1}{r|}{235K} & 13.44 & 9477.47 & 5093.05 & \textbf{998.54} \\
\multicolumn{1}{l|}{cycle2} & 160K & \multicolumn{1}{r|}{16K} & 0.94 & 12.34 & \textbf{8.32} & 66.76 \\
\multicolumn{1}{l|}{cycle3} & 272K & \multicolumn{1}{r|}{4.1K} & 0.84 & 17.69 & \textbf{10.94} & 341.39
\end{tabular}
\end{table*}

Figure ~\ref{fig:nfa-type-time} shows the execution time of the baseline, \od and \pmatch  algorithms for EPL and Email, also noting the number of temporal matches.  The BGP, which is in common for all executions in this experiment, returns 47K matches on EPL and 862K matches on Email.  When the temporal constraint is applied, the number of matches is reduced, and is presented on the $x$-axis.  For example, $\ta_1$ returns 1.8K
 matches on EPL and 36K on Email.  

We observe that \pmatch is the most efficient algorithm for all queries and both datasets, returning in under 0.12 sec for EPL, and in under 17 sec for Email in all cases.  The \od algorithm outperforms the baseline in all cases, but is slower than \pmatch.  The performance difference between the baseline and \od is due to a join between two large relations in baseline, compared to multiple joins over smaller relations in \od. 

We also observe that the relative performance of the algorithms depends on the number of matches, and explore this relationship further in the next experiment.  To
compare algorithm performance across different BGPs and
datasets, we use the timed automaton $\ta_0$ of
Figure~\ref{fig:inc-nfa}, which generalizes $\ta_1$ from $2$ to $m$ edges.
$\ta_0$ specifies that edges in a matching should appear
repeatedly in a strict temporal order.
We use $\ta_0$ (with $m = 2$, 3,
or 4 as appropriate) as the temporal constraint for paths of length 2 and 3, and for cycles of length 2, 3, 4, for three of the datasets. Because the Contact dataset is a bipartite graph, we used it for in-star ($x_2 \xrightarrow {y_1} x_1 \xleftarrow{y_2} x_3 $) and out-star($x_2 \xleftarrow {y_1} x_1 \xrightarrow{y_2} x_3 $) BGPs of size 2 and 3.

Table~\ref{tab:diff-pattern} summarizes the results.  It shows number of BGP matchings (``BGP''), number of matchings accepted by $\ta_0$ (``BGP+TA''),  and running times of computing the BGP match only (``match''), and of computing both BGP and temporal matches using to the baseline, \od, and \pmatch algorithms. 

We observe that, for acylic patterns (\eg paths, i-star, o-star), \pmatch is
significantly faster than \od and baseline. For such patterns,
partial matchings are shared by many total matchings and by
larger partial matchings, benefiting the running time.
Interestingly, for cycles of size 2 and 3, \od is fastest,
followed by baseline. However, for cycles of size 4
\pmatch is once again the fastest algorithm.  The reason for this
is that there are far fewer cycles than possible partial
matchings, and in smaller cycles this causes \pmatch to run
slower. As cycle size increases, performance of \pmatch becomes
comparable to, or better, than of the other two algorithms.
Another graph characteristic that can affect \pmatch performance
is graph density, which we discuss in the next section. (Our
machine's RAM could not fit cycle4 for FB-Wall, so we did not conduct
that experiment.)


\subsection{Comparison to in-memory databases}
\label{sec:experiments:sql-db}

In this set of experiments, we compare the running time of temporal BGP matching with equivalent relational queries.  We used three datasets (EPL, Contact and Email) and queried them with cyclic and acyclic BGPs of size 2, with temporal constraints specified by 9 timed automata: $TA_e$ specifies an existential constraint (Figure~\ref{fig:tae}), while $TA_1 \ldots TA_8$ express constraints that are not existential. 

%

We showed SQL queries QTA1 and QTA7 in the introduction.  In the
Supplementary Materials we give a complete listing of the SQL queries, with
\insql{Path2}, \insql{Cycl2} and \insql{Star2} expressing the
BGPs used in our experiments, and \insql{QTAE}, \insql{QTA1}--\insql{QTA8}
implementing the temporal constraints. 
Most of these queries are quite complicated compared to their equivalent timed automata.

%
%
For \duck and \hyper, we loaded the relations Node, Edge and Active into memory. To improve performance of \duck, we defined indexes on \insql{Edge(src)}, \insql{Edge(dst)}, and \insql{Active(eid,time)}. To the best of our knowledge, \hyper does not support indexes. 

Table~\ref{table:query-ta} shows the execution time for each
query, comparing the running time of the best method based on
timed automata (\od or \pmatch, column ``TAA'') with \duck and \hyper. Observe that our algorithms are significantly faster for $\ta_1$, $\ta_2$, $\ta_5$ and $\ta_7$ for all 3 datasets, and have comparable performance to the best-performing relational system for $\ta_3$ and $\ta_4$. Relational systems outperform our algorithms on $\ta_e$ and $\ta_6$. For $\ta_e$, relational databases compute all possible matchings at all the time points in one shot, and then filter out those that fail the temporal constraint, which can be faster than an iterative process. Similarly, for $\ta_6$, the XOR operator can be implemented as a join-antijoin. Interestingly, for $\ta_8$ (set containment join, see Introduction) \duck is most efficient, followed by our algorithms, and then by \hyper. For the majority of other cases, \duck either ran out of memory (\omem in Table~\ref{table:query-ta}) or was the slowest system.  Results for EPL are in supplementary materials.
Note that our system was able to handle all queries within the allocated memory, with \duck and \hyper both ran out of memory in some cases. 


\subsection{Impact of graph properties on performance}
\label{sec:experiments:density}

\newcommand{\STAB}[1]{\begin{tabular}{@{}c@{}}#1\end{tabular}}
\begin{table}[!t]
\small 
\caption{Best-performing timed-automaton algorithm (TAA) compared to \duck and \hyper, for SQL queries in Figure~\ref{fig:query-ta}.}
\label{table:query-ta}
\begin{tabular}{c|lrrrr}
\multicolumn{2}{c}{}  & \multicolumn{3}{r}{\textbf{time (sec)}}  \\ \hline 
\multirow{9}{*}{\STAB{\rotatebox[origin=c]{90}{\textbf{EPL Path2}}}} 
& TA & \multicolumn{1}{r|}{\# matches} &  TAA & \duck & \hyper\\ \hline
&$\ta_e$ & \multicolumn{1}{r|}{35866} & \textbf{1.09} & 1.11 & 1.44 \\
&$\ta_1$ & \multicolumn{1}{r|}{1801} & \textbf{0.06} & 60.14 & 17.84 \\
&$\ta_2$ & \multicolumn{1}{r|}{374} & \textbf{0.05} & 69.89 & 11.36  \\
&$\ta_3$ & \multicolumn{1}{r|}{21035} & \textbf{0.04} & 0.07 & 0.13 \\
&$\ta_4$ & \multicolumn{1}{r|}{29726} & \textbf{0.06} & 0.07 & 0.13 \\
&$\ta_5$ & \multicolumn{1}{r|}{1714} & \textbf{0.07} & 18.57 & 0.39 \\
&$\ta_6$ & \multicolumn{1}{r|}{19578} & 0.54 & 0.12 & \textbf{0.09}  \\
&$\ta_7$ & \multicolumn{1}{r|}{257} & \textbf{1.3} & 8.22 & 5.56  \\
&$\ta_8$ & \multicolumn{1}{r|}{5377} & 0.23 & \textbf{0.17} & 0.21 \\ \hline
\multirow{9}{*}{\STAB{\rotatebox[origin=c]{90}{\textbf{EPL Cycle2}}}}
&$\ta_e$ & \multicolumn{1}{r|}{933} & 0.04 & 0.597 & \textbf{0.01} \\
&$\ta_1$ & \multicolumn{1}{r|}{35} & \textbf{0.01} & 2.67 & 0.41 \\
&$\ta_2$ & \multicolumn{1}{r|}{22} & \textbf{0.01} & 3.33 & 0.26 \\
&$\ta_3$ & \multicolumn{1}{r|}{418} & \textbf{0.01} & 0.1 & \textbf{0.01} \\
&$\ta_4$ & \multicolumn{1}{r|}{740} & \textbf{0.01} & 0.1 & \textbf{0.01} \\
&$\ta_5$ & \multicolumn{1}{r|}{90} & \textbf{0.02} & 0.53 & 1.19 \\
&$\ta_6$ & \multicolumn{1}{r|}{312} & 0.07 & 0.07 & \textbf{0.01} \\
&$\ta_7$ & \multicolumn{1}{r|}{0} & \textbf{0.05} & 1.18 & 5.15 \\
&$\ta_8$ & \multicolumn{1}{r|}{188} & 0.02 & \textbf{0.01} & 0.09 \\ \hline
\multirow{9}{*}{\STAB{\rotatebox[origin=c]{90}{\textbf{Contact O-Star2}}}}
& $\ta_e$ & \multicolumn{1}{r|}{169410} & \textbf{8.87} & 22.43 & 19.29 \\
& $\ta_1$ & \multicolumn{1}{r|}{0} & \textbf{0.67} & \omem & 316.5 \\
& $\ta_2$ & \multicolumn{1}{r|}{0} & \textbf{0.72} & \omem & 205.95 \\
& $\ta_3$ & \multicolumn{1}{r|}{101268} & \textbf{0.59} & 0.95 & 0.69 \\
& $\ta_4$ & \multicolumn{1}{r|}{107650} & \textbf{0.68} & 0.91 & \textbf{0.68} \\
& $\ta_5$ & \multicolumn{1}{r|}{4154} & \textbf{0.72} & \omem & 2.035 \\
& $\ta_6$ & \multicolumn{1}{r|}{10832} & 4.68 & 0.99 & \textbf{0.54} \\
& $\ta_7$ & \multicolumn{1}{r|}{76} & \textbf{15.12} & \omem & 102.89 \\
& $\ta_8$ & \multicolumn{1}{r|}{4646} & 1.3 & \textbf{1.1} & 1.16 \\ \hline
\multirow{9}{*}{\STAB{\rotatebox[origin=c]{90}{\textbf{Email Path2}}}}
& $\ta_e$ & \multicolumn{1}{r|}{719609} & 501.07 & 649.69 & \textbf{239.82} \\
& $\ta_1$ & \multicolumn{1}{r|}{35594} & \textbf{22.35} & \omem & \omem \\
& $\ta_2$ & \multicolumn{1}{r|}{801} & \textbf{3.73} & \omem & 1748.85 \\
& $\ta_3$ & \multicolumn{1}{r|}{443431} & \textbf{7.78} & 17.84 & 8.59 \\
& $\ta_4$ & \multicolumn{1}{r|}{455977} & \textbf{7.82} & 17.78 & 8.26 \\
& $\ta_5$ & \multicolumn{1}{r|}{2474} & \textbf{1.27} & \omem & 48.87 \\
& $\ta_6$ & \multicolumn{1}{r|}{693956} & 320.04 & 8.98 & \textbf{5.61} \\
& $\ta_7$ & \multicolumn{1}{r|}{390} & \textbf{327.33} & \omem & \omem \\
& $\ta_8$ & \multicolumn{1}{r|}{12919} & 19.32 & \textbf{8.02} & 21.58 \\ \hline
\multirow{9}{*}{\STAB{\rotatebox[origin=c]{90}{\textbf{Email Cycle2}}}}
& $\ta_e$ & \multicolumn{1}{r|}{14188} & \textbf{4.71} & 254.7 & 15.3 \\
& $\ta_1$ & \multicolumn{1}{r|}{843} & \textbf{0.83} & \omem & 1788.87 \\
& $\ta_2$ & \multicolumn{1}{r|}{240} & \textbf{0.64} & \omem & 1733.05 \\
& $\ta_3$ & \multicolumn{1}{r|}{6333} & 0.46 & \textbf{0.38} & 0.68 \\
& $\ta_4$ & \multicolumn{1}{r|}{11396} & 0.44 & \textbf{0.38} & 0.68 \\
& $\ta_5$ & \multicolumn{1}{r|}{1800} & \textbf{0.97} & 206.26 & 1.19 \\
& $\ta_6$ & \multicolumn{1}{r|}{4230} & 24.09 & 8.63 & \textbf{5.48} \\
& $\ta_7$ & \multicolumn{1}{r|}{134} & \textbf{7.1} & \omem & \omem \\
& $\ta_8$ & \multicolumn{1}{r|}{3441} & 1.79 & \textbf{0.219} & 49.79 \\ \hline 
\end{tabular}
\end{table}

In this set of experiments, we explore the effect of structural
density and temporal domain size.  We synthetically generated a
complete graph with 50 nodes and 2450 edges (the same size as the
EPL dataset)  and temporal density of $0.5$. We then sampled edges to create a graph with different structural densities. We use 4 representative BGPs: paths of length 3, and cycles of length 3. As temporal constraint we use $\ta_4$ from Figure~\ref{fig:all-nfa}, as it has low early rejection rate, thus serving as a worst case.

\begin{figure*}[t!]
\centering
\subfloat[path 3]{
\includegraphics[width = 0.45\columnwidth]{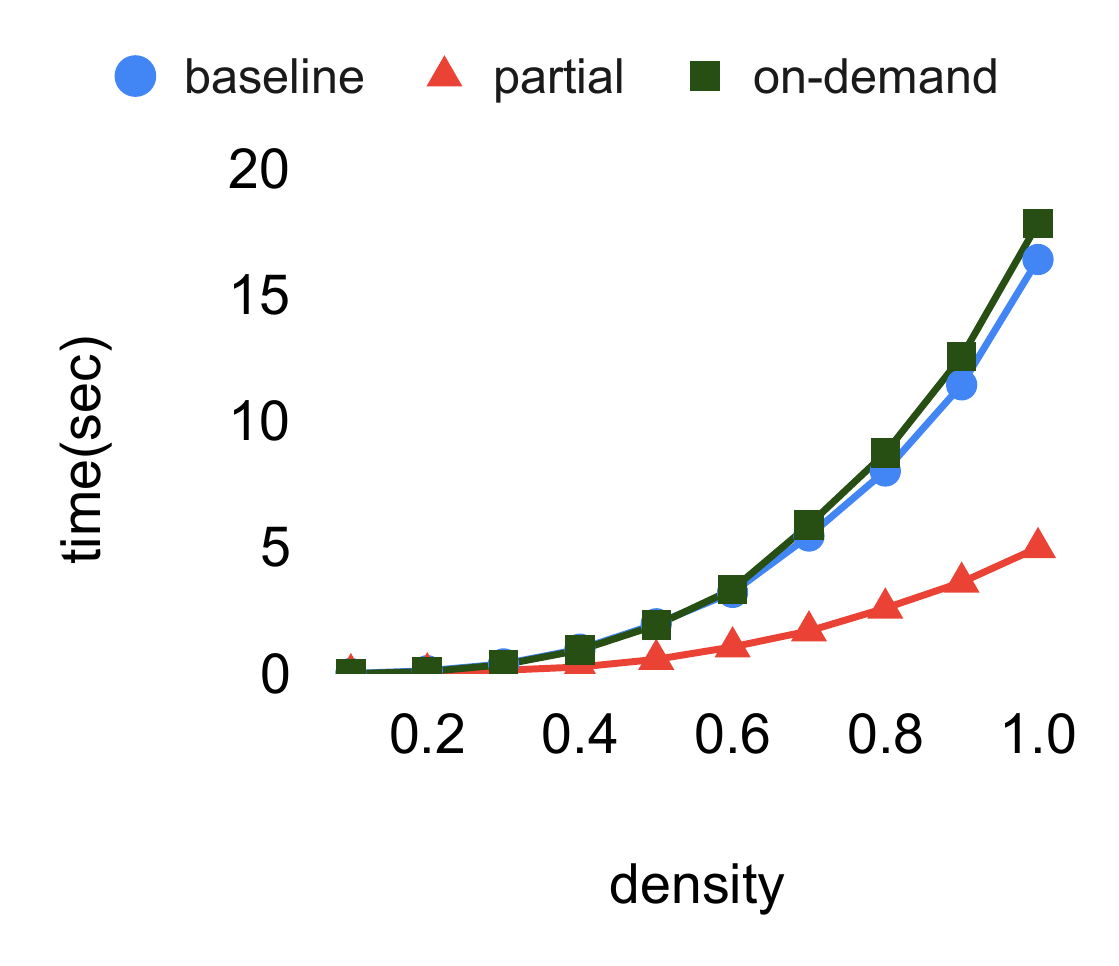}}
\subfloat[path 4]{
\includegraphics[width = 0.45\columnwidth]{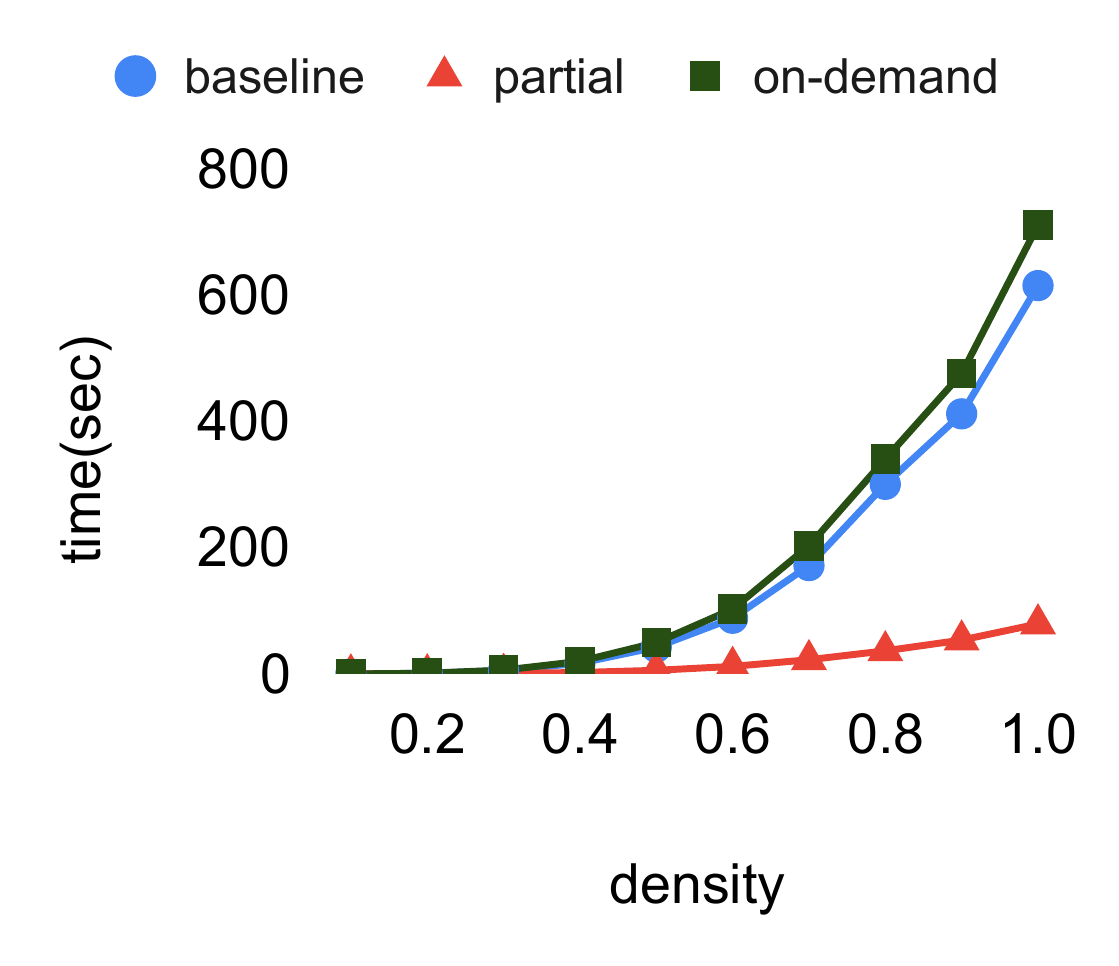}}
 \subfloat[cycle 3]{
 \includegraphics[width = 0.45\columnwidth]{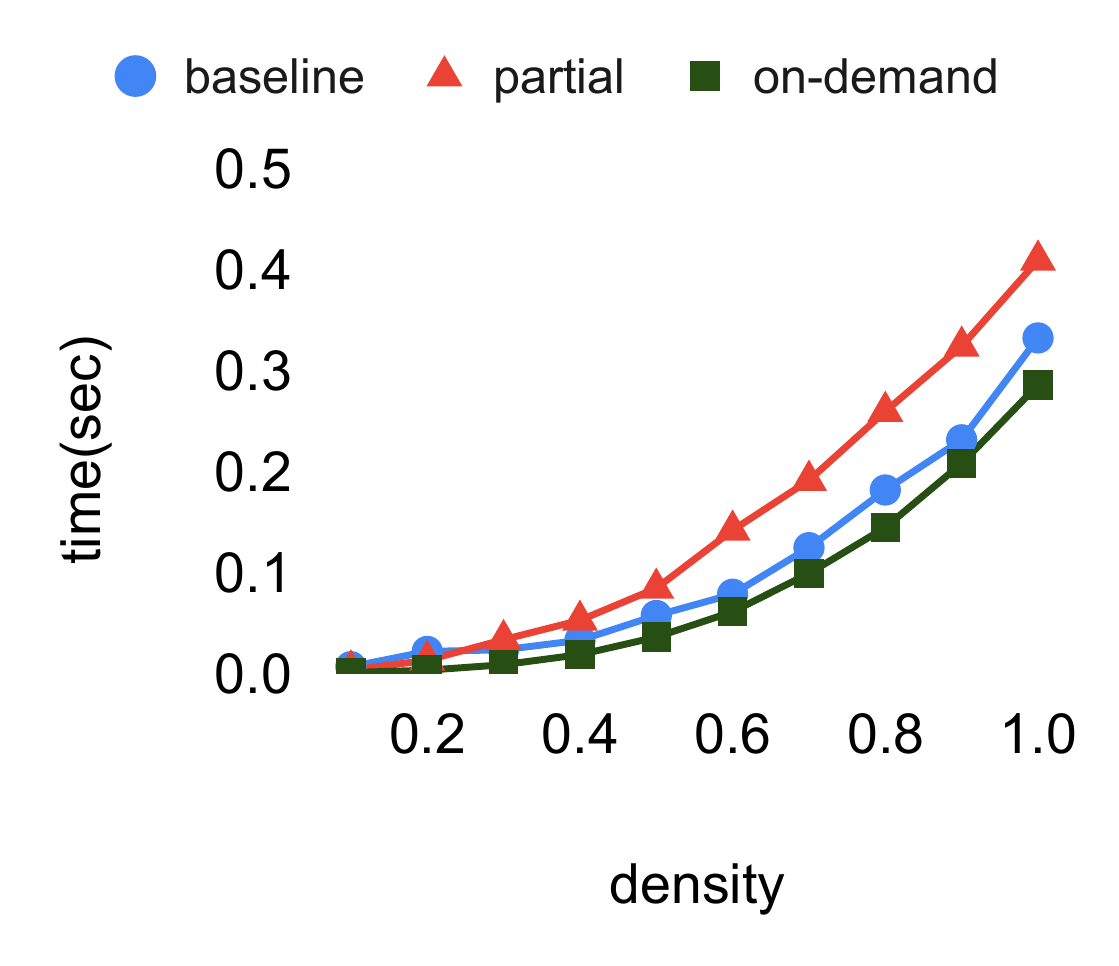}}
\subfloat[cycle 4]{%
 \includegraphics[width = 0.45\columnwidth]{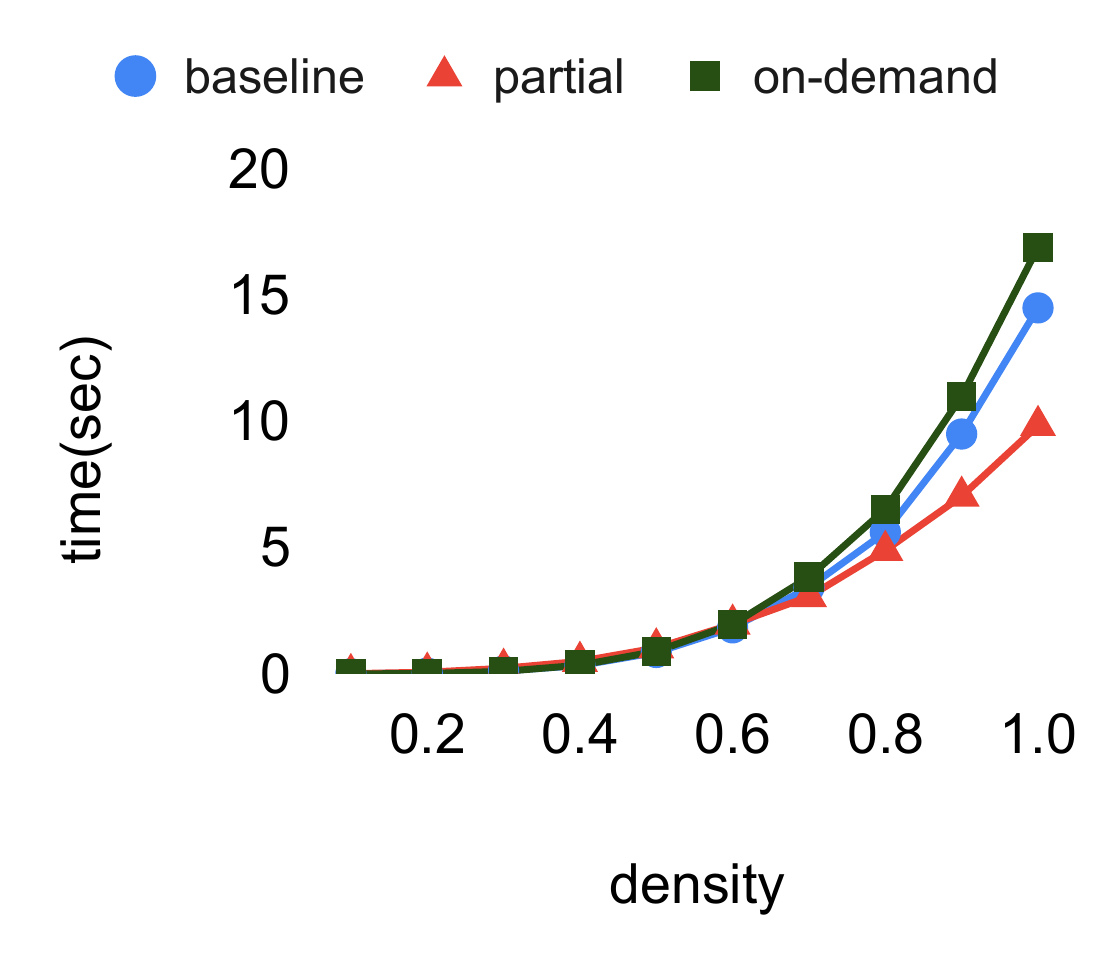}}
\caption {Running time as a function of structural density for 4 common BGPs, with timed automaton $\ta_4$ in Figure~\ref{fig:all-nfa}. 
}
\label{fig:graph-dense-result}
\end{figure*}



Figure~\ref{fig:graph-dense-result} shows the execution time of
each algorithm as a function of graph density, varying from 0.1
to 1.0, where density 1.0 corresponds to a complete graph. Observe that \pmatch outperforms \od for paths, particularly as graph density increases. For the cycle of size 3, baseline and
\od have better performance than \pmatch, but the performance gap decreases with increasing graph density.  Our experiments for BGPs: path of length 4 and cycle of length 4 show similar trend.
Notably, performance of \od is very close to the baseline.  


Next, we consider the impact of temporal domain size on performance.  In general, we expect execution times to increase with increasing temporal domain size.   
To measure this effect without changing the structure or the size of the graph, we synthetically changed the temporal resolution of the Email dataset, creating graphs with between 25 and 800 snapshots, and thus keeping the number of BGP matchings fixed. 

\begin{figure}[t!]
\centering
\subfloat[path2 with $\ta_1$ on Email]{
 \label{fig:temp-size}
\includegraphics[width = 0.50\columnwidth]{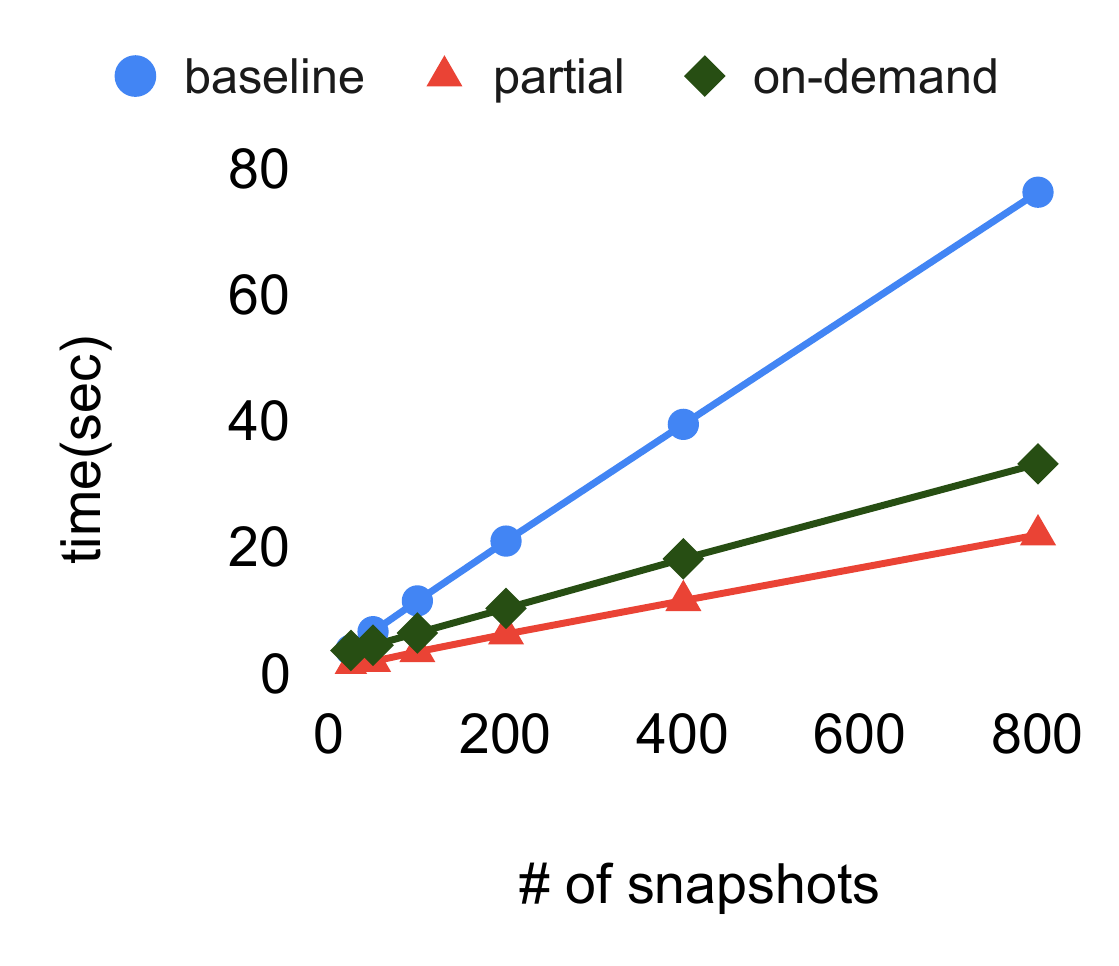}}
\subfloat[path2 with $\ta_2$ on Email]{
 \label{fig:selectivity}
\includegraphics[width = 0.50\columnwidth]{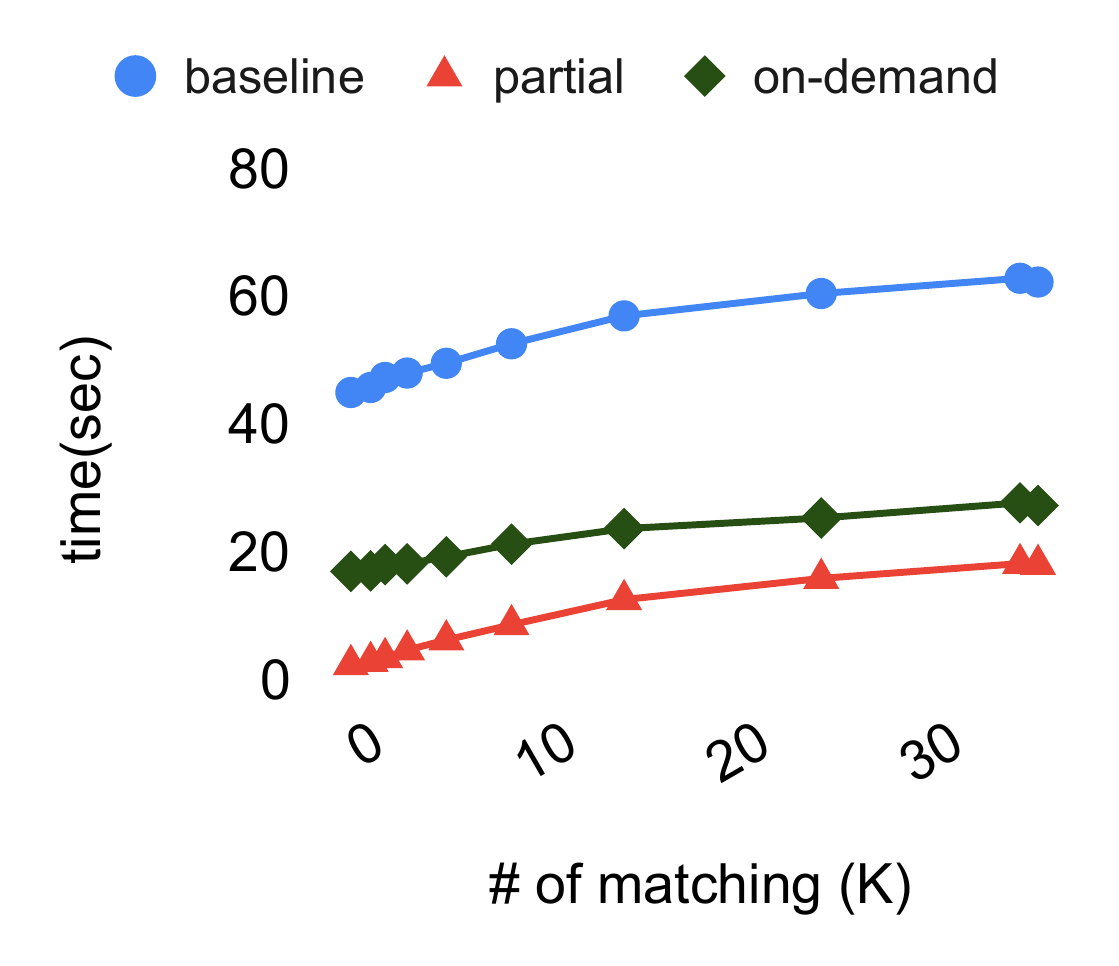}}
\caption{(a) running time vs. temporal domain size; (b) relationship between running time and result set size.}
\label{fig:temp-size-selectivity}
\end{figure}

Figure~\ref{fig:temp-size} shows the result of executing temporal BGP with path of length 2 and time automaton $\ta_1$ (Figure~\ref{fig:bgp-cycle2:ta1}).  Observe that the execution time of all algorithms increases linearly, with \pmatch scaling best with increasing number of snapshots. 




Finally, we study the relationship between result set size of a temporal BGP and algorithm performance. For this, we executed the path of length 2 BGP on the Email dataset, with timed automaton  $\ta_2$ in Figure~\ref{fig:bgp-cycle2:ta2}, and manipulated selectivity by varying the clock condition from $c < 0$ to $c < 1024$ on the logarithmic scale. With these settings, the temporal BGP accepts between 0 and 36K matchings.  Figure~\ref{fig:selectivity} presents the result of this experiment, showing the running time (in sec) on the $x$-axis and the number of temporal BGP matchings (in thousands) on the $y$-axis. Observe that the running time increases linearly with increasing number of accepted matchings for all algorithms, and the slope of increase is small.

\subsection{Impact of the number of clocks and automaton size on performance}
\label{sec:experiments:automata}

In our final set of experiments, we investigate the impact of automaton size and of the number of clocks on performance, while keeping all other parameters fixed to the extent possible. To do this, we fix the BGP and vary the size of the automaton, as follows.  We  fix the BGP to cycle4 and take $\ta_0$ (Figure~\ref{fig:inc-nfa}) with $m=4$ as a starting point. We can unfold the cycle of states, thus doubling the number of states but resulting in an equivalent automaton. We do this doubling seven times, until we obtain 256 states.

\begin{figure}[t!]
\centering
\subfloat[\# states]{%
\label{fig:nfa-size}%
\includegraphics[width = 0.45\columnwidth]{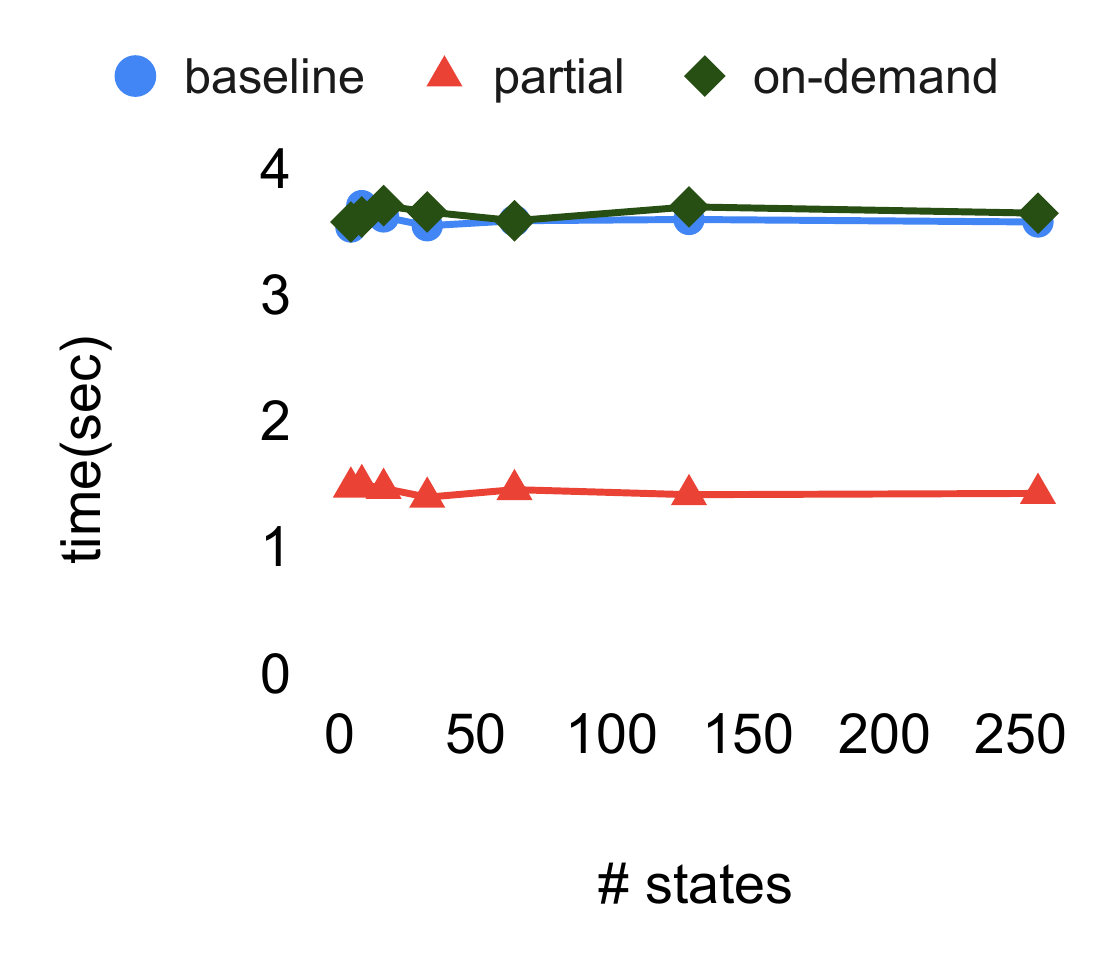}}
\subfloat[\# clocks]{%
\label{fig:clock}%
\includegraphics[width = 0.45\columnwidth]{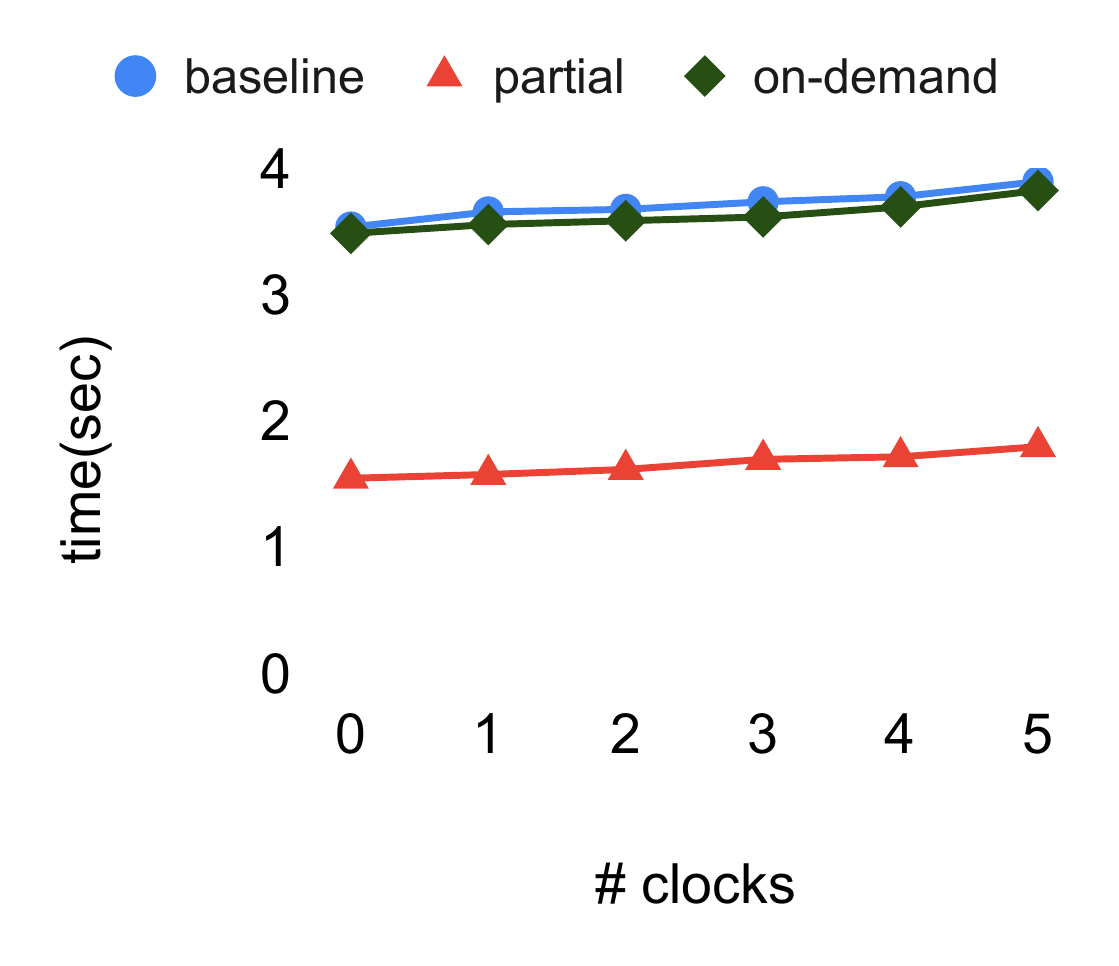}}
\caption{Running time as a function of automaton size for cycle of size 4 with $\ta_0$  (Figure~\ref{fig:inc-nfa}) on EPL.}
\label{fig:nfa-size-clock}
\vspace{-0.5cm}
\end{figure}

Figure~\ref{fig:nfa-size} shows the execution times on the EPL dataset. Observing that the execution times remain constant, we conclude that automaton size does not significantly impact performance.




Finally, to investigate the impact of the number of clocks on performance, we added multiple clocks to the automaton in Figure~\ref{fig:inc-nfa} and we reset  all clocks at every state transition. To ensure that any possible difference is not due to the change of  output size, the condition of each clock is set to $\mathit{true}$.  Figure~\ref{fig:clock} shows the result of this experiment, with the number of clocks on the $x$-axis and the execution time in seconds on the $y$-axis. Observe that the running time of all algorithms increases very slightly with increasing number of clocks.  We thus conclude that the computational overhead of storing and updating clocks is low. 

 \section{ Related Work }
 \label{sec:related}
During the past several decades, researchers considered different aspects of graph pattern matching, see Conte~\etal~\cite{conte2004thirty} for a survey. 
The majority of temporal graph models use either time
points~\cite{DBLP:conf/sigmod/GurukarRR15,DBLP:journals/corr/abs-1107-5646,DBLP:conf/wsdm/ParanjapeBL17}
or
intervals~\cite{DBPL2017,xu2017time,DBLP:conf/edbt/ZufleREF18,rost2022distributed}
to enrich graphs with temporal information. In our work, we
associate each edge in a graph with a set of time points, which
is an appropriate representation when events --- such as messages
between users or citations --- are instantaneous and so do not
have a duration.  It is an interesting topic for further research
to investigate when and how an interval-based approach can
be encoded by a point-based approach.  This depends also on the
considered graph model, and the
considered class of queries and temporal constraints.

A prominent line of work where the point-based approach is
adopted is that of mining frequent temporal subgraphs,
called temporal
motifs~\cite{DBLP:conf/sigmod/GurukarRR15,DBLP:journals/corr/abs-1107-5646,DBLP:conf/wsdm/ParanjapeBL17}.
There, the focus is typically on existential temporal constrains, aiming
to identify graph patterns with a specific temporal order among
the edges, such as in our Example~\ref{ex:exist-const}.
Timed automata can easily specify such constrains. An
important type of a motif is a $\delta$ temporal
motif~\cite{DBLP:conf/wsdm/ParanjapeBL17}, where all the edges
occur inside the period of $\delta$ time units. Timed automata
can use one or multiple clocks to enforce such constrains.

Z{\"u}fle~\etal~\cite{DBLP:conf/edbt/ZufleREF18}  consider a particular class of temporal constraints where the time points within a query range are specified \emph{exactly} up to the translation of the query range into the temporal range of the graph. Such constraints are more general than existential constraints, in that they can represent gaps.  An interesting aspect of this work is that the history of each subgraph is represented as a string, and the temporal constraint is checked using substring search. While this method of expressing constraints can work over a set of time points, it is limited to ordered temporal constraints and does not support reoccurring edges.

There are various lines of research on querying temporal graphs 
that are complementary to our focus in this paper.  For example,
durable matchings~\cite{semertzidis2016durable,li2022durable} count the number of snapshots in which a matching exists.  Much attention has also
been paid to tracing unbounded paths in temporal graphs, under
various semantics, e.g., fastest, earliest arrival, latest
departure, time-forward, time travel, or continuous
\cite{DBLP:journals/tkde/ByunWK20,Debrouvier21,DBLP:conf/sigmod/JohnsonKLS16,DBLP:journals/pvldb/WuCHKLX14,DBLP:journals/tkde/WuCKHHW16,DBLP:conf/icde/WuHCLK16}.
A focus on unbounded paths is complementary to our work on patterns without path variables, but with powerful
temporal constraints.  Extending our framework with path
variables is an interesting direction for further research. 

An important aspect of pattern matching in graphs is
efficiently extracting the matchings. Early work
started with the back-tracking algorithm by Ullmann
\cite{ullmann1976algorithm}, with later improvements
\cite{vcibej2015improvements,ulman2011}.  Pruning strategies for
brute-force algorithms have been investigated as well \cite{DBLP:journals/pami/CarlettiFSV18,cordella1999performance,cordella2004sub}.
Approaches suitable for large graphs typically build up the set of
matchings in a relational table~\cite{lai2019distributed} by a
series of natural joins over the edge relation; the aim is then
to find an optimal join order. Until recently, the
best-performing approaches were based on edge-growing pairwise
join plans~\cite{DBLP:journals/pvldb/LaiQLC15,selinger1989access,DBLP:journals/pvldb/SunWWSL12}, but
a new family of vertex-growing plans, known as worst-case optimal
joins, have
emerged~\cite{arroyuelo2021worst,DBLP:journals/sigmod/NgoRR13,DBLP:journals/corr/abs-1210-0481},
with better performance for cyclic patterns such as triangles.
While we use the latter approach and implement our algorithms
using relational operators,  any method capable of finding
matchings on a static graph can be combined with our timed
automaton-based algorithms.

Another relevant direction is incremental graph pattern matching~\cite{amsj_subgraph,DBLP:conf/eurosys/BindschaedlerML21,DBLP:journals/tkde/ChenW10,fan2013incremental,DBLP:conf/sigmod/HanLL13,imran2022fast,DBLP:conf/sigmod/KimSHLHCSJ18,DBLP:conf/sigmod/KoLHLSSH21}, where the goal is to find and maintain patterns in an updating graph. 

\balance 
\section{Conclusions and Future Work}
\label{sec:conclusion}


In this paper, we proposed to use timed automata as a simple but powerful formalism for
specifying temporal constraints in temporal graph pattern matching.  We introduced algorithms that retrieve all temporal BGP matchings in a large graph, and presented results of an experimental evaluation, showing that this approach is practical, and identifying interesting performance trade-offs. Our code and data are available at \url{http://github.com/amirpouya/TABGP}.


An interesting open problem is how timed automata exactly compare to SQL in expressing temporal constraints (pinpointing the expressive power of SQL on ordered data is notoriously hard \cite{libkin_sql}).  It is also interesting to investigate the decidability and complexity of the containment problem for temporal BGPs based on timed automata.  
Another natural direction for further research is to adapt our framework to a temporal
graph setting where edges are active at durations (intervals),
rather than at separate timepoints.  Our hypothesis is that we can
encode any set of non-overlapping intervals by the set of
border-points.  We conjecture that timed automata under
such an encoding can express common constraints on intervals,
such as Allen's relations~\cite{allen1983maintaining}.

\bibliographystyle{spmpsci}       
\bibliography{temporal}

\begin{thebibliography}{10}
\providecommand{\url}[1]{{#1}}
\providecommand{\urlprefix}{URL }
\expandafter\ifx\csname urlstyle\endcsname\relax
  \providecommand{\doi}[1]{DOI~\discretionary{}{}{}#1}\else
  \providecommand{\doi}{DOI~\discretionary{}{}{}\begingroup
  \urlstyle{rm}\Url}\fi

\bibitem{allen1983maintaining}
Allen, J.F.: Maintaining knowledge about temporal intervals.
\newblock Communications of the ACM \textbf{26}(11), 832--843 (1983)

\bibitem{timed-automata}
Alur, R., Dill, D.: A theory of timed automata.
\newblock Theoretical Computer Science \textbf{126}, 183--235 (1994)

\bibitem{amsj_subgraph}
Ammar, K., McSherry, F., Salihoglu, S., Joglekar, M.: Distributed evaluation of
  subgraph queries using worst-case optimal low-memory dataflows.
\newblock PVLDB \textbf{11}(6), 691--704 (2018)

\bibitem{DBLP:journals/csur/AnglesABHRV17}
Angles, R., Arenas, M., Barcel{\'{o}}, P., Hogan, A., Reutter, J.L., Vrgoc, D.:
  Foundations of modern query languages for graph databases.
\newblock {ACM} Comput. Surv. \textbf{50}(5), 68:1--68:40 (2017).
\newblock \doi{10.1145/3104031}.
\newblock \urlprefix\url{http://doi.acm.org/10.1145/3104031}

\bibitem{arroyuelo2021worst}
Arroyuelo, D., Hogan, A., Navarro, G., Reutter, J.L., Rojas-Ledesma, J., Soto,
  A.: Worst-case optimal graph joins in almost no space.
\newblock In: SIGMOD (2021)

\bibitem{DBLP:conf/eurosys/BindschaedlerML21}
Bindschaedler, L., Malicevic, J., Lepers, B., Goel, A., Zwaenepoel, W.:
  Tesseract: distributed, general graph pattern mining on evolving graphs.
\newblock In: A.~Barbalace, P.~Bhatotia, L.~Alvisi, C.~Cadar (eds.) EuroSys
  '21: Sixteenth European Conference on Computer Systems, Online Event, United
  Kingdom, April 26-28, 2021, pp. 458--473. {ACM} (2021).
\newblock \doi{10.1145/3447786.3456253}.
\newblock \urlprefix\url{https://doi.org/10.1145/3447786.3456253}

\bibitem{mamoulis-set-containment-revisited}
Bouros, P., Mamoulis, N., et~al.: Set containment join revisited.
\newblock Knowledge and Information Systems \textbf{49}, 375--402 (2016)

\bibitem{timed-automata-survey}
Bouyer, P., Fahrenberg, U., K., G.L., Markey, N., Ouaknine, J., Worell, J.:
  Model checking real-time systems.
\newblock In: E.~Clarke, T.~Henzinger, H.~Veith, et~al. (eds.) Handbook of
  Model Checking, pp. 1001--1046. Springer (2018)

\bibitem{DBLP:journals/tkde/ByunWK20}
Byun, J., Woo, S., Kim, D.: Chronograph: Enabling temporal graph traversals for
  efficient information diffusion analysis over time.
\newblock {IEEE} Trans. Knowl. Data Eng. \textbf{32}(3), 424--437 (2020).
\newblock \doi{10.1109/TKDE.2019.2891565}.
\newblock \urlprefix\url{https://doi.org/10.1109/TKDE.2019.2891565}

\bibitem{carbone2015apache}
Carbone, P., Katsifodimos, A., Ewen, S., Markl, V., Haridi, S., Tzoumas, K.:
  Apache flink: Stream and batch processing in a single engine.
\newblock Bulletin of the IEEE Computer Society Technical Committee on Data
  Engineering \textbf{36}(4) (2015)

\bibitem{DBLP:journals/pami/CarlettiFSV18}
Carletti, V., Foggia, P., Saggese, A., Vento, M.: Challenging the time
  complexity of exact subgraph isomorphism for huge and dense graphs with
  {VF3}.
\newblock {IEEE} Trans. Pattern Anal. Mach. Intell. \textbf{40}(4), 804--818
  (2018).
\newblock \doi{10.1109/TPAMI.2017.2696940}.
\newblock \urlprefix\url{https://doi.org/10.1109/TPAMI.2017.2696940}

\bibitem{DBLP:journals/tkde/ChenW10}
Chen, L., Wang, C.: Continuous subgraph pattern search over certain and
  uncertain graph streams.
\newblock {IEEE} Trans. Knowl. Data Eng. \textbf{22}(8), 1093--1109 (2010).
\newblock \doi{10.1109/TKDE.2010.67}.
\newblock \urlprefix\url{https://doi.org/10.1109/TKDE.2010.67}

\bibitem{cheng2008fast}
Cheng, J., Yu, J.X., Ding, B., Philip, S.Y., Wang, H.: Fast graph pattern
  matching.
\newblock In: 2008 IEEE 24th International Conference on Data Engineering, pp.
  913--922. IEEE (2008)

\bibitem{vcibej2015improvements}
{\v{C}}ibej, U., Miheli{\v{c}}, J.: Improvements to ullmann's algorithm for the
  subgraph isomorphism problem.
\newblock International Journal of Pattern Recognition and Artificial
  Intelligence \textbf{29}(07), 1550,025 (2015)

\bibitem{conte2004thirty}
Conte, D., Foggia, P., Sansone, C., Vento, M.: Thirty years of graph matching
  in pattern recognition.
\newblock International journal of pattern recognition and artificial
  intelligence \textbf{18}(03), 265--298 (2004)

\bibitem{cordella1999performance}
Cordella, L.P., Foggia, P., Sansone, C., Vento, M.: Performance evaluation of
  the vf graph matching algorithm.
\newblock In: Proceedings 10th International Conference on Image Analysis and
  Processing, pp. 1172--1177. IEEE (1999)

\bibitem{cordella2004sub}
Cordella, L.P., Foggia, P., Sansone, C., Vento, M.: A (sub) graph isomorphism
  algorithm for matching large graphs.
\newblock IEEE transactions on pattern analysis and machine intelligence
  \textbf{26}(10), 1367--1372 (2004)

\bibitem{curley2016engsoccerdata}
Curley, J.: Engsoccerdata: English soccer data 1871-2106.
\newblock R Package Version 0.1 \textbf{5} (2016)

\bibitem{Debrouvier21}
Debrouvier, A., Parodi, E., Perazzo, M., Soliani, V., Vaisman, A.: A model and
  query language for temporal graph databases.
\newblock VLDB Journal \textbf{30}(5) (2021).
\newblock \doi{https://doi.org/10.1007/s00778-021-00675-4}

\bibitem{fan2010graph}
Fan, W., Li, J., Ma, S., Tang, N., Wu, Y., Wu, Y.: Graph pattern matching: from
  intractable to polynomial time.
\newblock Proceedings of the VLDB Endowment \textbf{3}(1-2), 264--275 (2010)

\bibitem{fan2013incremental}
Fan, W., Wang, X., Wu, Y.: Incremental graph pattern matching.
\newblock ACM Transactions on Database Systems (TODS) \textbf{38}(3), 1--47
  (2013)

\bibitem{stijn-cer}
Grez, A., Riveros, C., Ugarte, M., Vansummeren, S.: A formal framework for
  complex event recognition.
\newblock TODS \textbf{46}(4) (2021)

\bibitem{gupta1993maintaining}
Gupta, A., Mumick, I.S., Subrahmanian, V.S.: Maintaining views incrementally.
\newblock ACM SIGMOD Record \textbf{22}(2), 157--166 (1993)

\bibitem{DBLP:conf/sigmod/GurukarRR15}
Gurukar, S., Ranu, S., Ravindran, B.: {COMMIT:} {A} scalable approach to mining
  communication motifs from dynamic networks.
\newblock In: T.K. Sellis, S.B. Davidson, Z.G. Ives (eds.) Proceedings of the
  2015 {ACM} {SIGMOD} International Conference on Management of Data,
  Melbourne, Victoria, Australia, May 31 - June 4, 2015, pp. 475--489. {ACM}
  (2015).
\newblock \doi{10.1145/2723372.2737791}.
\newblock \urlprefix\url{https://doi.org/10.1145/2723372.2737791}

\bibitem{DBLP:conf/sigmod/HanLL13}
Han, W., Lee, J., Lee, J.: Turbo\({}_{\mbox{iso}}\): towards ultrafast and
  robust subgraph isomorphism search in large graph databases.
\newblock In: K.A. Ross, D.~Srivastava, D.~Papadias (eds.) Proceedings of the
  {ACM} {SIGMOD} International Conference on Management of Data, {SIGMOD} 2013,
  New York, NY, USA, June 22-27, 2013, pp. 337--348. {ACM} (2013).
\newblock \doi{10.1145/2463676.2465300}.
\newblock \urlprefix\url{https://doi.org/10.1145/2463676.2465300}

\bibitem{imran2022fast}
Imran, M., G{\'e}vay, G.E., Quian{\'e}-Ruiz, J.A., Markl, V.: Fast datalog
  evaluation for batch and stream graph processing.
\newblock World Wide Web  (2022)

\bibitem{DBLP:conf/sigmod/JohnsonKLS16}
Johnson, T., Kanza, Y., Lakshmanan, L.V.S., Shkapenyuk, V.: Nepal: a path query
  language for communication networks.
\newblock In: A.~Arora, S.~Roy, S.~Mehta (eds.) Proceedings of the 1st {ACM}
  {SIGMOD} Workshop on Network Data Analytics, NDA@SIGMOD 2016, San Francisco,
  California, USA, July 1, 2016, pp. 6:1--6:8. {ACM} (2016).
\newblock \doi{10.1145/2980523.2980530}.
\newblock \urlprefix\url{https://doi.org/10.1145/2980523.2980530}

\bibitem{DBLP:conf/sigmod/KimSHLHCSJ18}
Kim, K., Seo, I., Han, W., Lee, J., Hong, S., Chafi, H., Shin, H., Jeong, G.:
  Turboflux: {A} fast continuous subgraph matching system for streaming graph
  data.
\newblock In: G.~Das, C.M. Jermaine, P.A. Bernstein (eds.) Proceedings of the
  2018 International Conference on Management of Data, {SIGMOD} Conference
  2018, Houston, TX, USA, June 10-15, 2018, pp. 411--426. {ACM} (2018).
\newblock \doi{10.1145/3183713.3196917}.
\newblock \urlprefix\url{https://doi.org/10.1145/3183713.3196917}

\bibitem{DBLP:conf/sigmod/KoLHLSSH21}
Ko, S., Lee, T., Hong, K., Lee, W., Seo, I., Seo, J., Han, W.: iturbograph:
  Scaling and automating incremental graph analytics.
\newblock In: G.~Li, Z.~Li, S.~Idreos, D.~Srivastava (eds.) {SIGMOD} '21:
  International Conference on Management of Data, Virtual Event, China, June
  20-25, 2021, pp. 977--990. {ACM} (2021).
\newblock \doi{10.1145/3448016.3457243}.
\newblock \urlprefix\url{https://doi.org/10.1145/3448016.3457243}

\bibitem{DBLP:journals/corr/abs-1107-5646}
Kovanen, L., Karsai, M., Kaski, K., Kert{\'{e}}sz, J., Saram{\"{a}}ki, J.:
  Temporal motifs in time-dependent networks.
\newblock CoRR \textbf{abs/1107.5646} (2011).
\newblock \urlprefix\url{http://arxiv.org/abs/1107.5646}

\bibitem{DBLP:journals/pvldb/LaiQLC15}
Lai, L., Qin, L., Lin, X., Chang, L.: Scalable subgraph enumeration in
  mapreduce.
\newblock Proc. {VLDB} Endow. \textbf{8}(10), 974--985 (2015).
\newblock \doi{10.14778/2794367.2794368}.
\newblock \urlprefix\url{http://www.vldb.org/pvldb/vol8/p974-lai.pdf}

\bibitem{lai2019distributed}
Lai, L., Qing, Z., Yang, Z., Jin, X., Lai, Z., Wang, R., Hao, K., Lin, X., Qin,
  L., Zhang, W., et~al.: Distributed subgraph matching on timely dataflow.
\newblock Proceedings of the VLDB Endowment \textbf{12}(10), 1099--1112 (2019)

\bibitem{DBLP:journals/tkdd/LeskovecKF07}
Leskovec, J., Kleinberg, J.M., Faloutsos, C.: Graph evolution: Densification
  and shrinking diameters.
\newblock {TKDD} \textbf{1}(1), 2 (2007).
\newblock \doi{10.1145/1217299.1217301}.
\newblock \urlprefix\url{http://doi.acm.org/10.1145/1217299.1217301}

\bibitem{li2022durable}
Li, F., Zou, Z., Li, J.: Durable subgraph matching on temporal graphs.
\newblock IEEE Transactions on Knowledge and Data Engineering  (2022)

\bibitem{libkin_sql}
Libkin, L.: Expressive power of {SQL}.
\newblock Theoretical Computer Science \textbf{296}, 379--404 (2003)

\bibitem{mcsherry2013differential}
McSherry, F., Murray, D.G., Isaacs, R., Isard, M.: Differential dataflow.
\newblock In: CIDR (2013)

\bibitem{DBLP:conf/dbpl/MoffittS17}
Moffitt, V.Z., Stoyanovich, J.: Temporal graph algebra.
\newblock In: Proceedings of The 16th International Symposium on Database
  Programming Languages, {DBPL} 2017, Munich, Germany, September 1, 2017, pp.
  10:1--10:12 (2017).
\newblock \doi{10.1145/3122831.3122838}.
\newblock \urlprefix\url{http://doi.acm.org/10.1145/3122831.3122838}

\bibitem{DBPL2017}
Moffitt, V.Z., Stoyanovich, J.: Temporal graph algebra.
\newblock In: Proceedings of The 16th International Symposium on Database
  Programming Languages, {DBPL} 2017, Munich, Germany, September 1, 2017, pp.
  10:1--10:12 (2017).
\newblock \doi{10.1145/3122831.3122838}.
\newblock \urlprefix\url{http://doi.acm.org/10.1145/3122831.3122838}

\bibitem{murray2013naiad}
Murray, D.G., McSherry, F., Isaacs, R., Isard, M., Barham, P., Abadi, M.:
  Naiad: a timely dataflow system.
\newblock In: Proceedings of the Twenty-Fourth ACM Symposium on Operating
  Systems Principles, pp. 439--455 (2013)

\bibitem{DBLP:journals/pvldb/Neumann11}
Neumann, T.: Efficiently compiling efficient query plans for modern hardware.
\newblock Proc. {VLDB} Endow. \textbf{4}(9), 539--550 (2011).
\newblock \doi{10.14778/2002938.2002940}.
\newblock \urlprefix\url{http://www.vldb.org/pvldb/vol4/p539-neumann.pdf}

\bibitem{DBLP:conf/sigmod/0001MK15}
Neumann, T., M{\"{u}}hlbauer, T., Kemper, A.: Fast serializable multi-version
  concurrency control for main-memory database systems.
\newblock In: T.K. Sellis, S.B. Davidson, Z.G. Ives (eds.) Proceedings of the
  2015 {ACM} {SIGMOD} International Conference on Management of Data,
  Melbourne, Victoria, Australia, May 31 - June 4, 2015, pp. 677--689. {ACM}
  (2015).
\newblock \doi{10.1145/2723372.2749436}.
\newblock \urlprefix\url{https://doi.org/10.1145/2723372.2749436}

\bibitem{DBLP:journals/sigmod/NgoRR13}
Ngo, H.Q., R{\'{e}}, C., Rudra, A.: Skew strikes back: new developments in the
  theory of join algorithms.
\newblock {SIGMOD} Rec. \textbf{42}(4), 5--16 (2013).
\newblock \doi{10.1145/2590989.2590991}.
\newblock \urlprefix\url{https://doi.org/10.1145/2590989.2590991}

\bibitem{ojagh2021person}
Ojagh, S., Saeedi, S., Liang, S.H.: A person-to-person and person-to-place
  covid-19 contact tracing system based on ogc indoorgml.
\newblock ISPRS International Journal of Geo-Information \textbf{10}(1), 2
  (2021)

\bibitem{DBLP:conf/wsdm/ParanjapeBL17}
Paranjape, A., Benson, A.R., Leskovec, J.: Motifs in temporal networks.
\newblock In: M.~de~Rijke, M.~Shokouhi, A.~Tomkins, M.~Zhang (eds.) Proceedings
  of the Tenth {ACM} International Conference on Web Search and Data Mining,
  {WSDM} 2017, Cambridge, United Kingdom, February 6-10, 2017, pp. 601--610.
  {ACM} (2017).
\newblock \doi{10.1145/3018661.3018731}.
\newblock \urlprefix\url{https://doi.org/10.1145/3018661.3018731}

\bibitem{DBLP:conf/sigmod/RaasveldtM19}
Raasveldt, M., M{\"{u}}hleisen, H.: Duckdb: an embeddable analytical database.
\newblock In: P.A. Boncz, S.~Manegold, A.~Ailamaki, A.~Deshpande, T.~Kraska
  (eds.) Proceedings of the 2019 International Conference on Management of
  Data, {SIGMOD} Conference 2019, Amsterdam, The Netherlands, June 30 - July 5,
  2019, pp. 1981--1984. {ACM} (2019).
\newblock \doi{10.1145/3299869.3320212}.
\newblock \urlprefix\url{https://doi.org/10.1145/3299869.3320212}

\bibitem{reza2020approximate}
Reza, T., Ripeanu, M., Sanders, G., Pearce, R.: Approximate pattern matching in
  massive graphs with precision and recall guarantees.
\newblock In: Proceedings of the 2020 ACM SIGMOD International Conference on
  Management of Data, pp. 1115--1131 (2020)

\bibitem{rost2022distributed}
Rost, C., Gomez, K., T{\"a}schner, M., Fritzsche, P., Schons, L., Christ, L.,
  Adameit, T., Junghanns, M., Rahm, E.: Distributed temporal graph analytics
  with gradoop.
\newblock The VLDB Journal \textbf{31}(2), 375--401 (2022)

\bibitem{rust-itertools}
Rust-Itertools: rust-itertools/itertools.
\newblock \urlprefix\url{https://github.com/rust-itertools/itertools}

\bibitem{selinger1989access}
Selinger, P.G., Astrahan, M.M., Chamberlin, D.D., Lorie, R.A., Price, T.G.:
  Access path selection in a relational database management system.
\newblock In: Readings in Artificial Intelligence and Databases, pp. 511--522.
  Elsevier (1989)

\bibitem{semertzidis2016durable}
Semertzidis, K., Pitoura, E.: Durable graph pattern queries on historical
  graphs.
\newblock In: 2016 IEEE 32nd International Conference on Data Engineering
  (ICDE), pp. 541--552. IEEE (2016)

\bibitem{DBLP:journals/pvldb/SunWWSL12}
Sun, Z., Wang, H., Wang, H., Shao, B., Li, J.: Efficient subgraph matching on
  billion node graphs.
\newblock Proc. {VLDB} Endow. \textbf{5}(9), 788--799 (2012).
\newblock \doi{10.14778/2311906.2311907}.
\newblock
  \urlprefix\url{http://vldb.org/pvldb/vol5/p788\_zhaosun\_vldb2012.pdf}

\bibitem{ullmann1976algorithm}
Ullmann, J.R.: An algorithm for subgraph isomorphism.
\newblock Journal of the ACM (JACM) \textbf{23}(1), 31--42 (1976)

\bibitem{ulman2011}
Ullmann, J.R.: Bit-vector algorithms for binary constraint satisfaction and
  subgraph isomorphism.
\newblock Journal of Experimental Algorithmics \textbf{15}, 1.6:1--1.6:64
  (2011).
\newblock \doi{10.1145/1671970.1921702}

\bibitem{DBLP:journals/corr/abs-1210-0481}
Veldhuizen, T.L.: Leapfrog triejoin: a worst-case optimal join algorithm.
\newblock In: Proceedings 17th International Conference on Database Theory, pp.
  96--106 (2014)

\bibitem{viswanath-2009-activity}
Viswanath, B., Mislove, A., Cha, M., Gummadi, K.P.: On the evolution of user
  interaction in facebook.
\newblock In: Proceedings of the 2nd ACM SIGCOMM Workshop on Social Networks
  (WOSN'09) (2009)

\bibitem{wood_survey}
Wood, P.: Query languages for graph databases.
\newblock SIGMOD Record \textbf{41}(1), 50--60 (2012)

\bibitem{DBLP:journals/pvldb/WuCHKLX14}
Wu, H., Cheng, J., Huang, S., Ke, Y., Lu, Y., Xu, Y.: Path problems in temporal
  graphs.
\newblock Proc. {VLDB} Endow. \textbf{7}(9), 721--732 (2014).
\newblock \doi{10.14778/2732939.2732945}.
\newblock \urlprefix\url{http://www.vldb.org/pvldb/vol7/p721-wu.pdf}

\bibitem{DBLP:journals/tkde/WuCKHHW16}
Wu, H., Cheng, J., Ke, Y., Huang, S., Huang, Y., Wu, H.: Efficient algorithms
  for temporal path computation.
\newblock {IEEE} Trans. Knowl. Data Eng. \textbf{28}(11), 2927--2942 (2016).
\newblock \doi{10.1109/TKDE.2016.2594065}.
\newblock \urlprefix\url{https://doi.org/10.1109/TKDE.2016.2594065}

\bibitem{DBLP:conf/icde/WuHCLK16}
Wu, H., Huang, Y., Cheng, J., Li, J., Ke, Y.: Reachability and time-based path
  queries in temporal graphs.
\newblock In: 32nd {IEEE} International Conference on Data Engineering, {ICDE}
  2016, Helsinki, Finland, May 16-20, 2016, pp. 145--156. {IEEE} Computer
  Society (2016).
\newblock \doi{10.1109/ICDE.2016.7498236}.
\newblock \urlprefix\url{https://doi.org/10.1109/ICDE.2016.7498236}

\bibitem{xie2007survey}
Xie, J., Yang, J.: A survey of join processing in data streams.
\newblock In: Data Streams, pp. 209--236. Springer (2007)

\bibitem{xu2017time}
Xu, Y., Huang, J., Liu, A., Li, Z., Yin, H., Zhao, L.: Time-constrained graph
  pattern matching in a large temporal graph.
\newblock In: L.~Chen, C.S. Jensen, C.~Shahabi, X.~Yang, X.~Lian (eds.) Web and
  Big Data, pp. 100--115. Springer International Publishing, Cham (2017)

\bibitem{DBLP:journals/vldb/YangW03}
Yang, J., Widom, J.: Incremental computation and maintenance of temporal
  aggregates.
\newblock {VLDB} J. \textbf{12}(3), 262--283 (2003).
\newblock \doi{10.1007/s00778-003-0107-z}.
\newblock \urlprefix\url{https://doi.org/10.1007/s00778-003-0107-z}

\bibitem{yoo2003slurm}
Yoo, A.B., Jette, M.A., Grondona, M.: Slurm: Simple linux utility for resource
  management.
\newblock In: Workshop on job scheduling strategies for parallel processing,
  pp. 44--60. Springer (2003)

\bibitem{DBLP:journals/cacm/ZahariaXWDADMRV16}
Zaharia, M., Xin, R.S., Wendell, P., Das, T., Armbrust, M., Dave, A., Meng, X.,
  Rosen, J., Venkataraman, S., Franklin, M.J., Ghodsi, A., Gonzalez, J.,
  Shenker, S., Stoica, I.: Apache spark: a unified engine for big data
  processing.
\newblock Commun. {ACM} \textbf{59}(11), 56--65 (2016).
\newblock \doi{10.1145/2934664}.
\newblock \urlprefix\url{http://doi.acm.org/10.1145/2934664}

\bibitem{zhao2010Comm}
Zhao, Q., Tian, Y., He, Q., Oliver, N., Jin, R., Lee, W.C.: Communication
  motifs: A tool to characterize social communications.
\newblock In: Proceedings of the 19th ACM International Conference on
  Information and Knowledge Management, CIKM '10, p. 1645–1648. Association
  for Computing Machinery, New York, NY, USA (2010).
\newblock \doi{10.1145/1871437.1871694}.
\newblock \urlprefix\url{https://doi.org/10.1145/1871437.1871694}

\bibitem{zfy-temporal-clique}
Zhu, K., Fletcher, G., Yakovets, N.: Leveraging temporal and topological
  selectivities in temporal-clique subgraph query processing.
\newblock In: ICDE (2021)

\bibitem{DBLP:conf/edbt/ZufleREF18}
Z{\"{u}}fle, A., Renz, M., Emrich, T., Franzke, M.: Pattern search in temporal
  social networks.
\newblock In: M.H. B{\"{o}}hlen, R.~Pichler, N.~May, E.~Rahm, S.~Wu, K.~Hose
  (eds.) Proceedings of the 21st International Conference on Extending Database
  Technology, {EDBT} 2018, Vienna, Austria, March 26-29, 2018, pp. 289--300.
  OpenProceedings.org (2018).
\newblock \doi{10.5441/002/edbt.2018.26}.
\newblock \urlprefix\url{https://doi.org/10.5441/002/edbt.2018.26}

\end{thebibliography}

\appendix 
\newpage 
\section{Supplementary materials for Section~\ref{sec:experiments}}

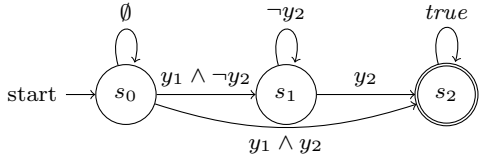
\begin{figure}[h!]
    \centering
    \begin{tikzpicture}[->, on grid, node distance = 60pt]
    \node[state, initial] (s0) {$s_0$};
    \node[state, right of = s0] (s1) {$s_1$};
    \node[state, right of = s1, accepting] (s2) {$s_2$};
    \draw (s0)
  (s0) edge[loop above] node{$\emptyset$} (s0)
  (s1) edge[loop above] node{$\neg y_2$} (s1)
  (s2) edge[loop above] node{$\mathit{true}$} (s2)

  (s0) edge[above] node{$y_1 \land  \neg y_2$} (s1)
  (s0) edge[below, bend right = 18] node{$y_1 \land y_2 $} (s2)
  (s1) edge[above] node{$y_2$} (s2);
    \end{tikzpicture}
    \caption{Version of $\ta_4$ for two edge variables.}
      \label{fig:apx:ta4-2}
\end{figure}


We present the SQL queries for the temporal BGPs with timed
automata used in the experiments.  For this experiment, we used BGPs with two edge variables. Thus, $\ta_4$, originally shown in Figure~\ref{fig:all-nfa}, is adapted here to two edge variables, as shown in Figure~\ref{fig:apx:ta4-2}.

\begin{lstlisting}
Path2: SELECT E1.eid AS y1, E2.eid AS y2 FROM edge E1, edge AS E2 WHERE E1.dst = E2.src
Cycl2: SELECT E1.eid AS y1, E2.eid AS y2 FROM edge E1, edge AS E2 WHERE E1.dst = E2.src and E2.dst = E1.src
Star2: SELECT E1.eid AS y1, E2.eid AS y2 FROM edge E1, edge AS E2 WHERE E1.dst = E2.dst
\end{lstlisting}

  \begin{lstlisting}
QTAE: SELECT  M.y1, M.y2 FROM matching M, active A1,active A2
      WHERE M.y1 = A1.eid AND M.y2 = A2.eidand A1.time < A2.time
  \end{lstlisting}
      
  \begin{lstlisting}
QTA1: WITH matching(y1,y2) as (...),
Succ AS
(SELECT y1, y2, A1.eid AS e1, A2.eid AS e2
 FROM matching, active A1, active A2
 WHERE (A1.eid = y1 OR A1.eid = y2) AND (A2.eid = y1 OR A2.eid = y2)
   AND A1.time<A2.time AND NOT EXISTS
         (SELECT * FROM active A3
          WHERE (A3.eid = y1 or A3.eid = y2)
            AND A1.time<A3.time AND A3.time<A2.time))
SELECT * FROM matching M
WHERE NOT ( EXISTS (SELECT * FROM active WHERE eid = y2)
            AND NOT EXISTS (SELECT * FROM active WHERE eid = y1) )
  AND ( NOT EXISTS (SELECT * FROM active A1, active A2
                    WHERE A1.eid = y1 AND A2.eid = y2)
        OR (SELECT MIN(time) FROM active WHERE eid = y1) <
           (SELECT MIN(time) FROM active WHERE eid = y2) )
  AND NOT EXISTS (SELECT * FROM Succ
                  WHERE M.y1 = y1 AND M.y2 = y2 AND e1 = e2);
  \end{lstlisting}

  \begin{lstlisting}
QT2: WITH matching(y1, y2) AS (...),
Succ AS 
(SELECT y1, y2, A1.eid as e1, A2.eid as e2, A1.time as t1, A2.time as t2
 FROM matching, active A1, active A2
 WHERE (A1.eid = y1 OR A1.eid = y2) AND (A2.eid = y1 OR A2.eid = y2)
   AND A1.time<A2.time AND NOT EXISTS
       (SELECT * FROM active A3
        WHERE (A3.eid = y1 or A3.eid = y2)
          AND A1.time<A3.time AND A3.time<A2.time))
SELECT *  FROM matching M
WHERE NOT ( EXISTS (SELECT * FROM active WHERE eid = y2)
            AND NOT EXISTS (SELECT * FROM active WHERE eid = y1) )
  AND ( NOT EXISTS (SELECT * FROM active A1, active A2
                    WHERE A1.eid = y1 AND A2.eid = y2)
        OR (SELECT MIN(time) FROM active WHERE eid = y1) <
           (SELECT MIN(time) FROM active WHERE eid = y2) )
  AND NOT EXISTS (SELECT * FROM Succ
                  WHERE M.y1 = y1 AND (e1 = e2  OR t2 - t1 >= 3)
  \end{lstlisting}

  \begin{lstlisting}
QTA3: WITH matching(y1, y2) AS (...)
SELECT M.y1, M.y2 FROM Matching M, active A1, active A2
WHERE M.y1 = A1.eid AND M.y2 = A2.eid
GROUP BY M.y1, M.y2 HAVING MIN(A1.time) < MIN(A2.time)
  \end{lstlisting}

  \begin{lstlisting}
QTA4: WITH matching(y1, y2) AS (...)
SELECT M.y1, M.y2 FROM Matching M, active A1, active A2
WHERE M.y1 = A1.eid AND M.y2 = A2.eid
GROUP BY M.y1, M.y2 HAVING MIN(A1.time) <= MIN(A2.time)
\end{lstlisting}  

\begin{lstlisting}                  
QTA5: WITH matching(y1, y2) AS (...)
SELECT * FROM Matching WHERE NOT EXISTS
 (SELECT * FROM active WHERE 
      (time IN (SELECT time FROM active WHERE eid = y1)
       OR time IN (SELECT time FROM active WHERE eid = y2))
  AND (time NOT IN (SELECT time FROM active WHERE eid = y1)
       OR time NOT IN (SELECT time FROM active WHERE eid = y2)))
\end{lstlisting}

\begin{lstlisting}
QTA6: WITH matching(y1, y2) AS (...)
SELECT * FROM matching WHERE NOT EXISTS
 (SELECT * FROM active A1, active A2
  WHERE A1.eid = y1 AND A2.eid = y2 AND A1.time = A2.time)
\end{lstlisting}

\begin{lstlisting}
QTA7: WITH matching AS (...)
SELECT DISTINCT y1, y2
FROM matching, active A1, active A2, active B1, active B2
WHERE A1.eid = y1 AND A2.eid = y2 AND A1.time = A2.time
  AND B1.eid = y1 AND B2.eid = y2 AND B1.time = B2.time
  AND B1.time - A1.time > 3
  AND NOT EXISTS
      (SELECT * FROM active C
       WHERE A1.time < C.time AND C.time < B1.time
         AND NOT EXISTS (SELECT * FROM active C1, active C2
                         WHERE C1.time = C.time AND C2.time = C.time
                         AND C1.eid = y1 AND C2.eid = y2));
\end{lstlisting}

\begin{lstlisting}
QTA8: WITH matching(y1, y2) AS (...)
SELECT * FROM Matching WHERE NOT EXISTS
    (SELECT * FROM active WHERE eid = y1 AND time NOT IN
        (SELECT time FROM active WHERE eid = y2))
\end{lstlisting}

\end{document}